%% file: main.tex
\documentclass[10pt,journal,compsoc]{IEEEtran}
\ifCLASSOPTIONcompsoc
  \usepackage[nocompress]{cite}
\else
  \usepackage{cite}
\fi

\usepackage[tight,footnotesize]{subfigure}
\usepackage[algo2e, ruled, linesnumbered]{algorithm2e}
\usepackage{cite}
\usepackage{amsmath,amssymb,amsfonts}
\usepackage{algorithmic}
\usepackage{epsfig}
\usepackage{graphicx}
\usepackage{textcomp}
\usepackage{xcolor}
\usepackage{comment}
\usepackage{diagbox}
\usepackage[spaces,hyphens]{url}
\def\BibTeX{{\rm B\kern-.05em{\sc i\kern-.025em b}\kern-.08em
    T\kern-.1667em\lower.7ex\hbox{E}\kern-.125emX}}

\newtheorem{definition}{Definition}

\usepackage{multirow}

\begin{document}
\newcommand*{\sysname}{vContact}
\title{\sysname{}: Private WiFi-based IoT \\
  Contact Tracing 
  with Virus Lifespan}

\author{Guanyao Li,
        Siyan Hu,
        Shuhan Zhong, Wai Lun Tsui,
        and
        S.-H.~Gary Chan
       
\IEEEcompsocitemizethanks{\IEEEcompsocthanksitem  The authors are with the Department
of Computer Science and Engineering, The Hong Kong University of Science and Technology, Kowloon,
Hong Kong, China. \protect\\
E-mail: gliaw@cse.ust.hk, siyanhu@ust.hk, szhongaj@cse.ust.hk, ptsuiwl@ust.hk, and gchan@cse.ust.hk.
\IEEEcompsocthanksitem This work has been submitted to the IEEE for possible publication.
Copyright may be transferred without notice, after which this version may
no longer be accessible.
}
}


\IEEEtitleabstractindextext{%
\begin{abstract}
Covid-19
is primarily spread through contact with the virus which may survive on surfaces with a lifespan of hours or even days if not sanitized.
To curb its spread, it is hence
of vital importance to detect those who have been in contact with the virus for a sustained period of time, the so-called {\em close contacts}.
Most of the existing digital approaches for contact tracing focus only on direct face-to-face contacts. There has been little work on detecting indirect environmental contact,
which is to detect people coming into a contaminated area with the live virus, i.e., an area 
last visited by an infected person within the virus lifespan.

In this work, we study, for the first time, automatic IoT contact tracing when the virus has a lifespan which may depend on the disinfection frequency at a location. Leveraging the ubiquity of WiFi signals, we propose \sysname{},  a novel, private, pervasive and fully distributed WiFi-based IoT contact tracing approach.
Users carrying an IoT device (phone, wearable, dongle, etc.)
continuously scan WiFi access points (APs) and store their hashed IDs.  Given a confirmed case, the signals
are then uploaded to a server for other users to match in their local IoT devices for virus exposure notification.  \sysname{} is not based on device pairing, and no information of other users is stored locally.  The confirmed case does not need to have the device for it to work properly.
As WiFi data are sampled sporadically and asynchronously, \sysname{} uses 
novel and effective signal processing approaches and a 
similarity metric to align and match signals at any time.
We conduct extensive indoor and outdoor experiments to validate \sysname{} performance. Our results demonstrate that \sysname{} is effective and accurate for contact detection. The precision, recall and F1-score of contact detection are high (up to 90\%)
for close contact proximity ($2$m). Its performance is robust against AP numbers, AP changes and phone heterogeneity. Having implemented \sysname{} as an Android SDK and installed it on phones and smart watches, we present a case study to demonstrate the validity and implementability of our design in notifying its users about their exposure to the virus
with a specific lifespan.
\end{abstract}

\begin{IEEEkeywords}
Contact Tracing, Exposure Notification, COVID-19, Data Management and Analytics, Social Impacts.
\end{IEEEkeywords}}

\maketitle

\IEEEdisplaynontitleabstractindextext

%
\IEEEpeerreviewmaketitle

\input{1_introduction}

\input{2_related_work}

\input{4_contact_detection}

\input{6_evalution}

\input{7_case_study}
\input{9_conclusion}

\ifCLASSOPTIONcaptionsoff
  \newpage
\fi



%

%

\bibliographystyle{IEEEtran}
\bibliography{main}

\end{document}

%% file: 1_introduction.tex
\section{Introduction}
\label{introduction}

The outbreak of COVID-19 has had a profound impact on our lives and global economy.
COVID-19, like many other infectious diseases, is primarily spread through  viral contact.  Recent studies have shown that
the virus has a lifespan: in airborne droplets it can last more than 10 minutes, and on surfaces it
can survive for hours to days if not properly disinfected (in low temperatures it may last even longer)~\cite{link_covid}~\cite{link_bbc}.
The health of any person coming into contact with the live virus for a sustained period of time, say 15--30 minutes, may be at risk \cite{troncoso2020decentralized}.    
In order to effectively contain the spread of the disease,
tracing and quarantining these {\em close contacts} as soon as possible
is of paramount importance.

Traditionally, close contacts are traced manually through personal interviews with infected people by medical officers. Such a manual approach is labour-intensive and slow. Due to mis-memory, the contact information may be incomplete or error prone.  Furthermore, the patient may not know
everyone in his/her proximity, and
those coming into the same area within the virus lifespan after he/she has left.

To overcome the above limitations of manual tracing, we propose \sysname{}, a novel, private and digital contact tracing solution using the Internet of things (IoT)  with possibly location-dependent virus lifespan.  Anyone in contact with the living virus 
is considered at risk.  This includes those simultaneously located with the patient, and those sharing the same environment 
which the patient has left.
\sysname{} leverages
ubiquitous
WiFi signals to achieve pervasive, fully distributed and automated contact tracing.
Note that, although for concreteness our discussion will focus on WiFi signals, \sysname{} can be straightforwardly extended to other radio-frequency (RF) signals such as Bluetooth and their combinations.
To the best of our knowledge,  this is the first work on using  RF signals for  private IoT contact tracing with a virus lifespan.

\begin{figure}[!t]
  \centering
  \includegraphics[width=.99\linewidth]{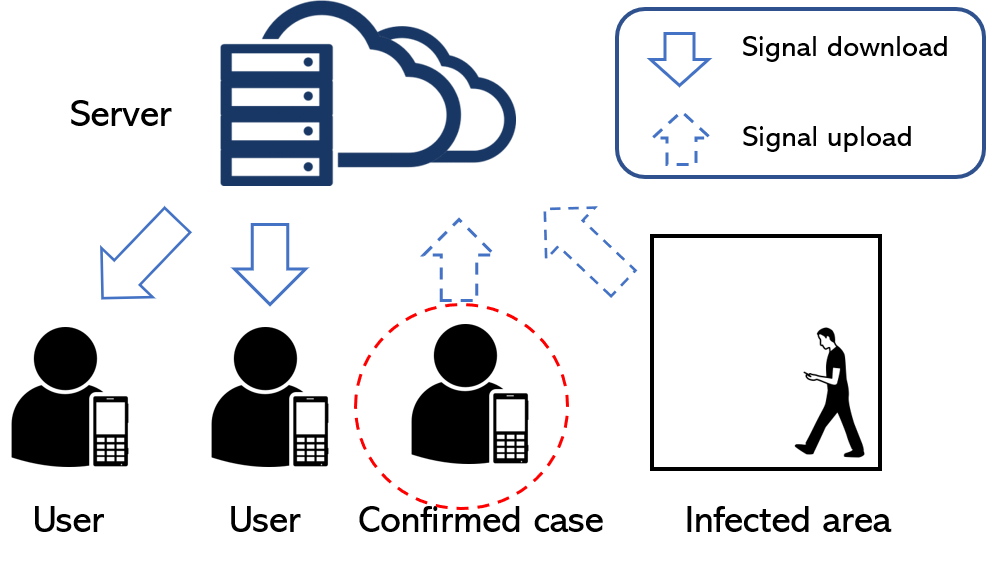}
  \caption{The process of \sysname{} contact tracing using WiFi. }
  \label{fig:data_collection}
\end{figure}

We illustrate the process of \sysname{} in Figure~\ref{fig:data_collection}.  
A user carries a WiFi-enabled IoT device (phone, wearable, dongle, etc.).
For concreteness and ease of illustration, we use a phone as the example in the figure.
With an installed app, it periodically scans for WiFi, with each scan collecting
a {\em signal vector} consisting of two elements: 1) the
signal IDs, which are the hashed (and optionally encrypted) values of the MAC addresses of
the WiFi access points~(APs); and 2) the corresponding received signal strength
indicators~(RSSIs) of the signal IDs. Each signal vector is associated with a timestamp, which is the
scanning/collection time of the signals. 
Over time, the device collects and stores a time series of the signal vectors, termed the
{\em signal profile}. The signal profile may be kept for a certain duration corresponding to the virus incubation period, usually 14 to 28 days for Covid-19.  

Upon positive confirmation in hospital, the patient
has the following two possibilities:
\begin{itemize}
\item {\em With the installed app:}
  With the consent of the patient, the health officer may access his/her signal profile. (Before
sharing with the officer,
the patient may blank out some time spans of the signal profile for personal reasons.)
Because the signal IDs are hashed (and possibly encrypted) from AP MACs, the officer does not know the patient's geo-locations, but only clusters of anonymized IDs and their collection times.  Based on that, the officer 
works with the patient to identify the physical venues presenting potential health risks to the public.
The corresponding anonymized IDs  are extracted and labelled
with their projected virus lifespan at the location at that time (depending on disinfection frequency).  The resultant
signal profile is then uploaded to a secure server for other IoT users to download and match with their own local profile in a distributed manner.  Upon detecting a close contact, the user is alerted at once in private to check their health condition and seek medical advice. 

\item {\em Without the app:} In this case, the confirmed case has to rely on his/her memory to recall the major venues and visit time as in the manual case.  Then some staff will go to these venues (the infected areas) to collect offline their WiFi information and label them with the visit time and viral lifespan at that time.  These manually labelled 
  data are then uploaded after being processed, and matched by the other IoT users the same way as in the previous case.
\end{itemize}

There have been prior works on automatic digital contact tracing.  Some use GPS~\cite{berke2020assessing} and cellular signals\cite{wesolowski2014commentary}.
While effective, these approaches do not work well in indoor environments due to signal blockage. Because they are also based on explicit user geo-locations, and such locations may be computed or stored in other's networks, the systems raise concerns regarding location privacy.  Due to these reasons, they have not become mainstream.
Some privacy-preserving approaches based on phone-to-phone pairing using
Bluetooth Low Energy (BLE) have attracted much attention and been implemented recently~\cite{troncoso2020decentralized}\cite{bell2020tracesecure}\cite{chan2020pact}.
However, they work for only direct face-to-face contact tracing, and cannot be applied for
environmental contamination, i.e., 
the case with non-zero spatial-temporal virus lifespan. 

\sysname{}  complements the above and
may be  integrated with them
(such as~\cite{troncoso2020decentralized}\cite{chan2020pact}).
Compared with prior arts, \sysname{} has the following strengths and unique features:

\begin{itemize}
\item {\em Contact detection with virus lifespan:} \sysname{} is the first piece of work for ubiquitous IoT contact tracing to capture the realistic scenario of virus lifespan which may be location-dependent and temporally varied depending on disinfection operation.  It comprehensively covers, in a single framework, those in direct face-to-face contact  {\em and} indirect environmental exposure in the areas previously visited by an infected person. The lifespan of the virus, set at the time of signal upload, may be heterogeneous and customized depending on the frequency of disinfection operation in the venue.
  
\item {\em No device-to-device pairing and communication:} 
  Prior contact tracing proposals based on Bluetooth require device pairing, which means both devices, including the infected one, have to have installed the app or software
in order  to work properly.
  To achieve  tracing effectiveness, they hence demand a high adoption rate (in the range of reportedly 40\% -- 70\%).  Moreover, such a device pairing approach  may suffer from replay/relay attack~\cite{ahmed2020survey} and raise privacy and security concerns~\cite{sapiezynski2017inferring}.
In contrast to such pairing, each \sysname{} device operates independently without any pairing or communication, and does not require the confirmed case to have already had the IoT device.  This greatly relaxes the adoption barrier and provides a graceful adoption path. Furthermore, users do not store any information of or exchange any messages with other users; it hence offers much better protection of user anonymity, privacy and attacks.

\item {\em Privacy by design:} \sysname{} is privacy-by-design.
%
First, it does not require a user registration process, and hence accesses no personal information such as names, phone numbers, IDs, contact lists, images/videos, etc.
%
%
Second, the collected data never leave the local storage without the explicit consent of the owner, and even so (i.e., the case of a confirmed case) no personally identifiable data are uploaded, and the data remain anonymous at the server.
Finally, \sysname{} is fully decentralized. The collected data are exclusively stored in one's own device, and the contact is computed and detected locally on the device in a scalable manner without any other centralized entity (party or server) having full information.
As no user data are stored anywhere beyond one's device, a user may exit the system at any time by device removal or app un-installation without leaving his/her data behind. 
Upon detection of close contact, \sysname{} conveys the message to its users in private. It is clear that such data fragmentation and minimization protect data privacy, and prevent data re-purposing, abuse, and mis-use.
Due to its distributed and hence scalable nature, it is deployable from small local communities to across a country.

\item {\em No GPS-based geo-location:
  } \sysname{} is not based on GPS.  It is based on the hashed values of WiFi MAC addresses (namely signal IDs) without storing 
  the user's physical geo-location.
  This leads to much stronger location confidentiality than other GPS-based  approaches, because the association of signal IDs to their physical locations takes an enormously and prohibitively large amount of manual work (that is to visit every indoor and outdoor spot of the city and logging down the locations of all the MAC addresses encountered).  Furthermore, unlike other GPS approaches, \sysname{} can detect indoor contacts and hence is more pervasive. 
  
\end{itemize}  

Detecting close contact using WiFi data is a challenging problem.
It is because signal vectors are sampled sporadically at random discrete times.
Such independence and asynchrony among IoT devices results in difficulties detecting contact at any arbitrary time. Furthermore,
signals may be sparsely sampled in the space (once every minute or so)
Therefore, the scanned IDs and their RSSIs at a location at a distinct time may be different because of the change in the environment.
Moreover, due to the device heterogeneity on antenna design and sensitivity,
the collected signal IDs may also differ for different users.

\begin{figure}
  \centering
  \includegraphics[width=.99\linewidth]{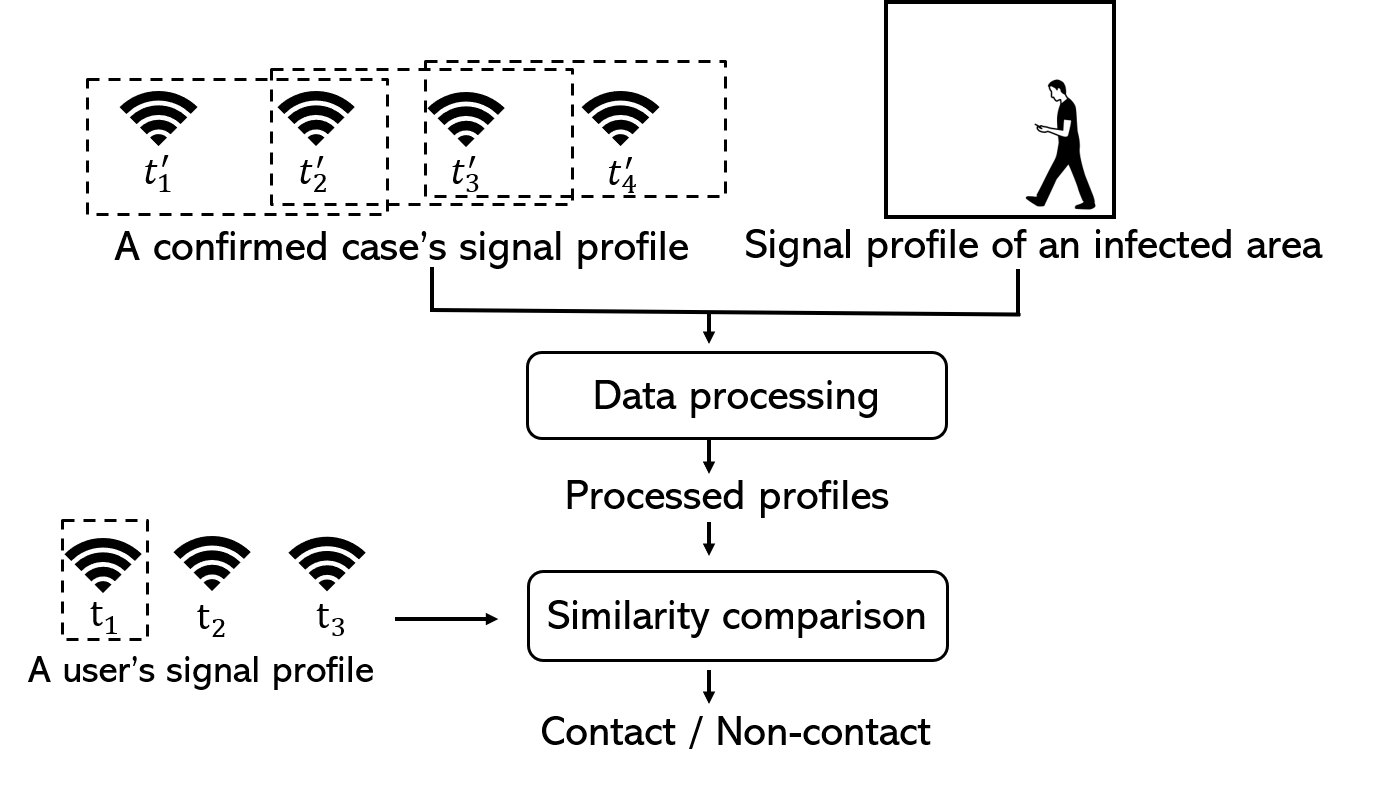}
  \caption{Overview of contact detection using WiFi in \sysname{}.}
  \label{fig:overview}
\end{figure}

\sysname{}
overcomes these problems by
employing an efficient approach to represent the values between consecutive signal vectors and a novel similarity metric to match signal values for contact tracing.
We present in Figure~\ref{fig:overview} an overview of \sysname{}.
It first processes the discrete signal profile from a confirmed case or infected area 
by transforming it
into a continuous profile (the processed profile).
Using that, given the signal vector of a user at time $t$, if $t$ falls in the time range of the virus lifespan of a processed vector, \sysname{} compares their level of matching using a novel signal similarity metric. If the similarity is larger than a given threshold, the user is said to be in contact with the virus at $t$. 
A user is identified as a close contact if the contact time exceeds a certain sustained period of time as specified by health officials.

\sysname{} is simple, and we have implemented it as
a software development kit (SDK)
and Android app.  We conduct extensive indoor and outdoor experiments with a diverse and representative set of IoT devices such as smart watches and phone brands in the market (Samsung, Honor,  Huawei  Nova,  Huawei  Mate30,  Xiaomi and OPPO).
Our results show that it 
achieves high precision, recall and F1-score~(up to $90\%$  for the contact proximity of $2$m), and its performance is robust against AP numbers, AP changes and devices of different brands.  \sysname{} can achieve good accuracy even when the AP number is low (as few as five APs in our experiments), meaning that it is widely applicable to city or suburban areas.

The remainder of this paper is organized as follows.
We introduce related works in 
Section~\ref{related_work}.
In Section~\ref{sec:contact_detection} we present the approach of \sysname{}.
We have implemented \sysname{} as an SDK, and
discuss the experiment setting and illustrative results  in Section~\ref{sec:evaluation}.  With the SDK, we have installed it in IoT smart watches and built 
an app, and present its implementation details and measurement in
Section~\ref{Sec:case study}. We conclude with future works in Section~\ref{sec:conclusion}.

%% file: 2_related_work.tex
\section{Related works}
\label{related_work}


Automatic contact tracing has attracted much attention in both academia and industry due to its importance in containing the the spread of the Covid-19 pandemic~\cite{ahmed2020survey}\cite{li2020covid}\cite{reichert2020survey}. In this section, we present the prior arts in the area.


Some studies have used signals which reveal user geo-location, such as GPS, cellular data, and radio frequency identification (RFID).
GPS signal provides a user's exact location for contact tracing~\cite{qi2013tracking}\cite{fitzsimons2020note}\cite{klopfenstein2020digital}\cite{reichert2020privacy}, but it is usually weak and noisy in indoor environments, limiting its contact coverage. Cellular data can be used to infer a user's public transportation trips~\cite{li2017estimating}\cite{li2017public}, which is crucial for contact tracing. Given the data, one can detect users taking the same bus, train, or subway with a confirmed case.
However, this approach often has high location errors, because the coverage of the cell tower is large, and close proximity is difficult to detect. 
Some researchers have also proposed using RFID to understand contact~\cite{isella2011close}\cite{salathe2010high}. However, special devices have to be deployed for data collection. Meanwhile, some geo-location based contact tracing systems have been deployed around the world, such as Corowarner in Turkey~\cite{Corowarner}, Aarogya Setu in India~\cite{ Aarogya_Setu}, Cotrack in Argentina~\cite{Cotrack}.
Corowarner and Aarogya Setu use GPS data, while Cotrack fuses signals of RFID, GPRS, GPS, and telecommunication technologies.
All the above works may be extended to contact tracing with the virus lifespan.  However, they may raise privacy concerns as they are based on user's physical geo-location.
By contrast,  \sysname{} offers much better location confidentiality, achieves better location accuracy, and is pervasive and easy to use. 

Location privacy is a major concern for contact tracing~\cite{cho2020contact}.
To better protect it, some works propose using a magnetometer~\cite{jeong2019smartphone}.
However, geomagnetism suffers from location ambiguity, which may lead to unsatisfactory proximity detection in practice.
There has also been much work based on
device-to-device message exchange using
Bluetooth~\cite{bell2020tracesecure}\cite{hekmati2020contain}\cite{sattler2020risk}\cite{xia2020return}.
User devices broadcast their ID using Bluetooth, and scan the nearby IDs. Based on the scanned IDs, one can know if he/she has had close contact with an infected case\cite{gunther2020tracing}.
Among the Bluetooth approaches, centralised solutions rely on a third-party server for contact tracing. 
Among these works, BlueTrace~\cite{bay2020bluetrace} and ROBERT~\cite{castelluccia2020robert} are the two most representative protocols. They use a decentralised framework to collect data, but a centralised system to analyze the exposure risks. Bluetooth data are collected via device-to-device communication and are stored locally. Once a user is infected, he/she can upload his/her scanned data to a security server for analysis. Users who are at risk will then be identified by the centralised system. The major difference of the two protocols is the way that people know their risks. In BlueTrace, the health authority would proactively contact the individuals who have a high likelihood of virus exposure, while users of ROBERT have to periodically probe the server for their risk score of exposure. Based on the BlueTrace protocol, the automatic contact tracing app TraceTogether~\cite{traceTogether} has been deployed in Singapore, which is the first national deployment of the Bluetooth-based contact tracing system. Based on a similar concept to that of TraceTogether, another system called COVIDSafe has been deployed in Australia to slow the spread of COVID-19~\cite{COVIDSafe}. Furthermore, DESIRE~\cite{castelluccia2020desire} is an extension of the ROBERT protocol, which is based on the same architecture of ROBERT with some major privacy improvements. 

Since a third-party server may raise the concern of possible data abuse, other works advocate a fully distributed approach, where the exposure detection and notifications are processed on an individual device. Representative works include PACT-UW~\cite{chan2020pact},  DP-3T~\cite{troncoso2020decentralized}, PACT-MIT~\cite{MIT_PACT} and Pronto-C2~\cite{avitabile2020towards}~(Note that both PACT-UW and PACT-MIT are termed as PACT in their origin papers). 
In these decentralised systems, users collect the encrypted IDs of their nearby users and store them locally. When someone is confirmed as being infected, he/she can upload his/her encrypted ID for other users to download for contact tracing. Compared with centralised solutions, only the encrypted IDs of infected cases are uploaded for the decentralised solutions, and contact information is distributed on user devices for storage. 

Based on the concept of decentralised systems, Google and Apple provide a toolkit for privacy-preserving contact tracing using Bluetooth~\cite{google_apple}. Some Bluetooth-based decentralised systems have also been deployed in some countries, such Covid Watch in the US~\cite{covid-watch} and SwissCovid~\cite{SwissCovid} in the Switzerland.
All these schemes are independently designed and very similar, apart from some minor variations in implementation and efficiency. 
All the above works focus on detecting face-to-face close contact, and they cannot be extended to the case with virus lifespan.  We propose a private WiFi-based approach to detect close contacts with {\em virus lifespan}. To the best of our knowledge, this is the first piece of work considering a virus lifespan in private contact tracing using WiFi. Moreover, no IoT device pairing or communication are needed in our proposed scheme, and hence no minimal adoption rate.

%% file: 4_contact_detection.tex
\section{
  \sysname{} Details}
\label{sec:contact_detection}

We present the details of \sysname{} in this section.
We first discuss its data processing approaches to construct the processed profile from the raw signal profile, for the patient with and without the installed app on their IoT devices
in Section~\ref{sec:data_process_app_users} and Section~\ref{sec:data_process_nonapp_users}, respectively. We then introduce
in Section~\ref{sec:signal_similarity_metric}
an efficient  and novel signal similarity metric to measure signal similarity, given a user's signal vector and a processed vector. We summarize \sysname{} and outline its contact detection algorithm in Section~\ref{sec:conatct_detection_algo}.

\input{3_Preliminary.tex}

\subsection{Profile processing for a patient with the app}
\label{sec:data_process_app_users}
Signals are not sampled continuously but at sporadic and random intervals.
Consequently, signal data are not continuously observable, leading to difficulty in comparing signal similarity at any arbitrary time.
We propose here a data processing approach to construct continuous
profiles from raw signal profiles for patients with our installed software.

\begin{figure}
  \centering
  \includegraphics[width=1\linewidth]{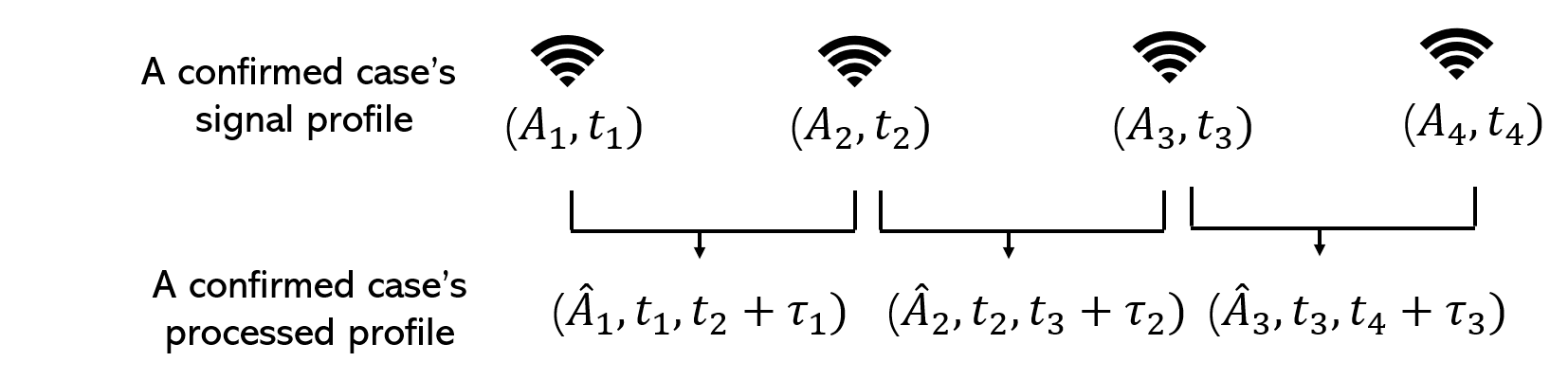}
  \caption{Signal profile processing for a confirmed case with app.}
  \label{fig:signal_processing_app_user}
\end{figure}

We show an example of signal profile processing in Figure~\ref{fig:signal_processing_app_user}. A confirmed case's signal profile $\{(A_1,t_1),(A_2,t_2),(A_3,t_3),(A_4,t_4)\}$ consists of some signal vectors at discrete times. We aim to construct a continuous processed profile from the raw signal profile so that the signal vector at any arbitrary time can be compared. To achieve this, we construct the processed vectors $\hat{A_i}$ from any two consecutive signal vectors $A_i$ and $A_{i+1}$, and consider the virus lifespan $\tau_i$. The virus lifespan $\tau_i$ may vary with time. 

\begin{definition}
(Processed vector) A processed vector is defined as: $\hat{A} = \{(a_1,s^{min}_1,s^{max}_1),(a_2,s^{min}_2,s^{max}_2),  ...,  (a_i, $ \\ $ s^{min}_i,s^{max}_i),...,(a_n,s^{min}_n,s^{max}_n)\}$, where $(a_i,s^{min}_i,s^{max}_i)$ denotes that the RSSI range of a signal $a_i$ is from $s^{min}_i$ to $s^{max}_i$.
\end{definition}

The signal strength in a processed vector is represented as a range instead of an exact value in a signal vector. Given two consecutive signal vectors in a signal profile $A_i = \{(a_1^i, s_1^i), ..., (a_j^i, s_j^i), ... (a_n^i, s_n^i)\}$ 
at $t_i$ and $A_{i+1} = \{(a_1^{i+1}, s_1^{i+1}), ..., (a_k^{i+1}, s_k^{i+1}), ... (a_m^{i+1}, s_m^{i+1})\}$ at $t_{i+1}$, the processed vector in the time range from $t_i$ to $t_{i+1}$ is denoted as $\hat{A} = \{(a_{\ell}, s^{min}_{\ell}, s^{max}_{\ell})|\ell = 1,2,...,|A_i \cup A_{i+1}|\}$, where $a_{\ell} \in A_i.a \cup A_{i+1}.a$ and $(s_{\ell}^{min}, s_{\ell}^{max})$ is denoted as
\begin{equation}
\left\{
             \begin{array}{lr}
            
             (\min(s_j^i,s_k^{i+1}), \max(s_j^i,s_k^{i+1})),& \mbox{for } a_{\ell} \in A_i \cap A_{i+1}; \\
             
             (\gamma, s_j^i) , & \mbox{for }  a_{\ell} \in A_{i}, a_{\ell}  \notin A_{i+1};   \\
            
             (\gamma, s_k^{i+1}) , &\mbox{for }  a_{\ell} \in A_{i+1}, a_{\ell} \notin  A_{i}.
            
             \end{array}
\right.
\end{equation}
Here, $\gamma$ is a value indicating a weak signal strength, which is set to be $-100$ in our experiments. 
Then, we construct a continuous processed profile from a confirmed case's signal profile considering the virus lifespan. We present a formal definition of a processed profile.

\begin{definition}
(Processed profile) A processed profile contains a sequence of processed vectors over time: $\hat{W} = \{(\hat{A}_1,t_1^{start}, t_1^{end}), (\hat{A}_2,t_2^{start}, t_2^{end}), ..., (\hat{A}_i,t_i^{start}, t_i^{end}), ...$ \\ $(\hat{A}_m,t_m^{start}, t_m^{end})\}$, where $\hat{A}_i$ is a processed vector for the time slot from $t_i^{start}$ to $t_i^{end}$, and $(t_i^{start}$, $t_i^{end})$ indicates the time slot of the virus lifespan.
\end{definition}

Given a confirmed case's signal profile $W = \{(A_1,t_1), (A_2,t_2), ..., (A_i,t_i), ... (A_n,t_n)\}$, the processed profile is represented as $\hat{W} = \{(\hat{A}_1,t_1, t_2 + \tau_1), (\hat{A}_2,t_2, t_3 + \tau_2), ..., (\hat{A}_i,t_i, t_{i+1} + \tau_i),..., (\hat{A}_{n-1},t_{n-1}, t_n + \tau_{n-1})\}$, where $\hat{A}_i$ is constructed from $A_i$ and $A_{i+1}$ and $\tau_i$ is the virus lifespan for the time slot from $t_i$ to $t_{i+1}$. Note that $\tau_i$ is given by the health officer, and it can vary for different time slots
depending on the frequency of disinfection operation in
the venues.

\subsection{Signal profile processing for infected areas}
\label{sec:data_process_nonapp_users}
For the case where the patient has not installed the app, we need to extract the signals in the infected areas through a survey (signal collection process).  
We can evaluate if a user has been in contact with an infected area by measuring the similarity of her/his signal vector and signal vectors of each position in the area.  However, collecting WiFi data for every position in the infected area is inefficient. We propose an efficient approach to construct the processed profile for an infected area using just some sampled signal data in the area. 

\begin{figure}
  \centering
  \includegraphics[width=1\linewidth]{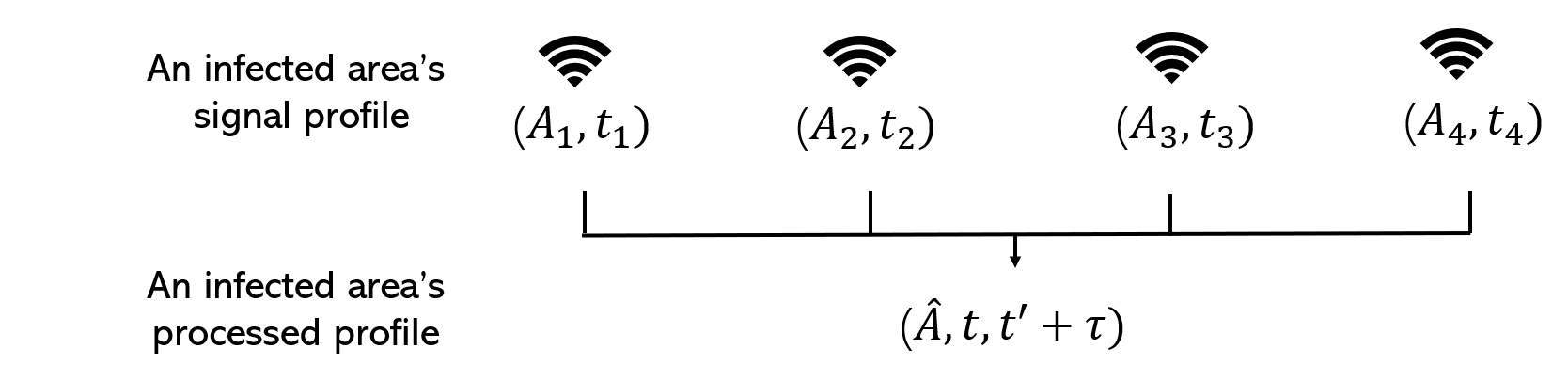}
  \caption{Signal profile processing for an infected area.}
  \label{fig:signal_processing_non_app}
\end{figure}

Instead of collecting signal data at every position, staff walk around the area with a WiFi-on device such as a phone or a Raspberry Pi. The collected signal profile is some signal vectors over time. To generate a representative processed profile for the area, we aggregate all signals and their RSSIs in the signal profile. As shown in Figure~\ref{fig:signal_processing_non_app}, we merge the signal vectors in the signal profile $\{(A_1, t_1),(A_2, t_2),(A_3, t_3),(A_4, t_4)\}$ which are collected in the infected area. We also consider the time range $[t,t']$ when a confirmed case stays in the area and the virus lifespan $\tau$ to construct the processed profile for the infected area.

The processed profile of an area is represented as $\hat{W} = (\hat{A}, t^{start},t^{end})$, where $\hat{A}$ is a processed vector and $[t^{start},t^{end}]$ is the time range of the virus lifespan. Given the signal profile collected in the infected area $W = \{(A_1, t_1),(A_2, t_2),...,(A_i, t_i),...(A_n, t_n)\}$, the time range of a confirmed case staying in the area $[t,t']$, and the virus lifespan $\tau$, the processed profile $\hat{W} = (\hat{A}, t^{start}, t^{end})$ is constructed as follows: 
$\hat{A} = \{( a_j, s_j^{min}, s_j^{max})|j = 1,2,...,|\cup_i^n A_i.a|\}$ where $a_j$ is a scanned signal in $W$~(i.e., $a_j \in \cup_i^n A_i.a$), and $s_j^{min}$ is the minimum signal strength of $a_j$ in $W$ while $s_j^{max}$ is the maximum signal strength of $a_j$ in $W$; the surviving time of the virus in the infected area is 
from $t$ to $t + \tau$.

\subsection{Signal similarity metric }
\label{sec:signal_similarity_metric}
We propose a signal similarity metric to compare the similarity of a signal vector with a processed vector for exposure detection. The metric considers the signal IDs overlap ratio and the RSSI difference.

\begin{table}
\caption{Average number of signals in a signal vector for various mobile phones}
\label{table: average number of signals}
\centering
\begin{tabular}{|c|c|c}
\hline
\textbf{Phone}         & \textbf{Average number of signals}  \\
\hline
Honor          & 75.00                                  \\
\hline
Huawei Mate 30 & 128.12                              \\
\hline
OPPO    & 180.16                             \\
\hline
Huawei Nova & 92.87\\
\hline
Xiaomi & 102.09 \\
\hline
\end{tabular}
\end{table}

Intuitively, the closer a user is to the location of the virus, the more  signal IDs are shared between the user's signal vectors and the vectors in the processed profile. Thus, we could use the overlap ratio of two vectors' signal IDs to indicate their proximity.
Given a user signal vector $A$ at time $t$ and a processed vector $\hat{A}$, the overlap ratio is calculated as:
\begin{equation}
O = \frac{|A.a \cap \hat{A}.a|}{\min(|A.a|, |\hat{A}.a|)},
\label{equ:signal_overlap}
\end{equation}
where $A.a$ is the Signal IDs in $A$, $\hat{A}.a$ is the signal IDs in $\hat{A}$, and $|\cdot|$ denotes the number of signal IDs.

The reason for using $\min(|A.a|,|\hat{A}.a|)$ is to mitigate the impact of the dynamic environment and device heterogeneity. An IoT device such as a phone or smart watch may scan different numbers of WiFi APs at a location at different times. 
Moreover, different IoT devices may have different abilities to scan signals, with two co-located devices scanning different numbers of signals. Table \ref{table: average number of signals} shows the average numbers of signals in a signal vector of various co-located phones in a shopping mall. The average number of signals is heterogeneous for different phones. The difference could be significant for some phones. In this case, using $|A.a|$, $|\hat{A}.a|$ or other terms~(e.g. $|A.a \cup \hat{A}.a|$) as the denominator will introduce more variance.

A signal could cover a large area, so it is possible that two vectors with a large proportion of common signals are not in close proximity. Thus, we also consider the RSSI difference to denote the proximity. If a user stays close with the virus, the RSSI difference of the same signal in two vectors should be small.
Given a user signal vector $A = \{(a_1,t_1), (a_2,t_2), ... (a_i,t_i), ... (a_n,t_n)\}$  and a processed vector $\hat{A} = \{(a_1,s_1^{min},s_2^{max}), (a_2,s_2^{min},s_2^{max}),  ...,(a_j,s_j^{min},\\$$s_j^{max}), ..., (a_m,s_m^{min},s_m^{max})\}$, for $a_k \in A.a \cap \hat{A}.a$, its RSSI difference is calculated as 
\begin{equation}
d(a_k) = \left\{
\begin{array}{lr}
     s^{min}_j - s_i , & s_i < s^{min}_j, \\
     s_i - s^{max}_j, & s_i > s^{max}_j,\\
     0, & \text{otherwise}.
\end{array}
\right.
\end{equation}

The average RSSI difference at a timestamp is defined as 
\begin{equation}
    D = \frac{\sum_{a_k \in ( A.a \cap \hat{A}.a)} d(a_k)}{|A.a \cap \hat{A}.a|},
\label{equ:signal_difference}
\end{equation}
where $|\cdot|$ denotes the number of signal IDs. 

When a user has contact with the virus, the overlap score $O$~(Equation \ref{equ:signal_overlap}) should be large, while the RSSI difference $D$~(Equation \ref{equ:signal_difference}) should be small. Therefore, we define the signal similarity of $A$ and $\hat{A}$ as 
\begin{equation}
\label{equ:proximity_comp}
    P(A, \hat{A}) = \frac{O}{D +1},
\end{equation}
where $0 \le P(A, \hat{A}) \le 1$. A larger $P(A, \hat{A})$ indicates closer proximity.

\subsection{The \sysname{} algorithm}
\label{sec:conatct_detection_algo}
\input{algorithm/indirect_contact_tracing_1}
Anyone having contact with the surviving virus may be at risk. Given a user's signal vector $A_i$ at $t_i$, if the timestamp $t_i$ is within the virus lifespan, and the similarity of $A_i$ and the processed profile of a confirmed case or an infected area is larger than a threshold, the user will be detected as having contact with the virus at $t_i$. The algorithm is presented in Algorithm \ref{algo_indirect_tracing}.


Given a user's signal profile $W_1$, the signal profile of a confirmed case or an infected area $W_2$, the virus lifespan $\{\tau_i|i = 1,2,...,|W_2| - 1\}$ and a proximity threshold $\alpha$, we first construct the processed profile from $W_2$ and $\{\tau_i|i = 1,2,...,|W_2| - 1\}$~(Line 4). Then, for each signal vector $A_i$ at time $t_i$ in $W_1$, if $t_i$ falls in the time slot of a processed vector in the processed profile, we calculate the signal similarity~(using Equation \ref{equ:proximity_comp}) at $t_i$~(Line 7 $\sim$ 9). If the similarity at $t_i$ is larger than the given threshold $\alpha$, the user is identified as having contact with the virus at $t_i$~(Line 11). The algorithm evaluates the similarity of each signal vector in $W_1$ and $\hat{W}$, and returns a list of detection results. The threshold $\alpha$ depends on how we define the contact proximity for close contact. We will discuss the relationship between the signal similarity and physical proximity, and the determination for the proximity threshold $\alpha$ in the following section.

%% file: 3_Preliminary.tex

We define 
signal vector and signal profile as follows.

\begin{definition}
(Signal vector) A signal vector $A$ is represented as $\{(a_1,s_1), (a_2,s_2), ..., (a_i,s_i), ..., (a_n,s_n)\}$, where $a_i$ is the signal ID (hashed and possibly encrypted AP MAC address) and $s_i$ is its RSSI. 
\end{definition}

\begin{definition}
(Signal profile) A user's signal profile is defined as a sequence of signal vectors over time: $W = \{(A_1,t_1), (A_2,t_2), ... , (A_i,t_i), ... ,(A_n, t_n)\}$, where $A_i$ is the signal vector scanned at time $t_i$. 
\end{definition}

In other words, a signal vector represents the signals and RSSIs scanned by an IoT device at a certain time, while a signal profile is a collection of the signal vectors over time.  The contact tracing is then stated as follows.
Given a user's signal profile $W = \{(A_1, t_1), (A_2, t_2),...,(A_i, t_i),...,(A_n, t_n)\}$, detect if the user has contact with the virus at each $t_i$ by comparing the similarity of the signal vector at that time with the signal profile of a confirmed case or an infected area.

%% file: algorithm/indirect_contact_tracing_1.tex
\SetArgSty{textnormal}
\let\oldnl\nl
\newcommand{\nonl}{\renewcommand{\nl}{\let\nl\oldnl}}
\begin{algorithm2e}
\SetAlgoLined

\textbf{Input}: A user's signal profile $W_1$ ;  \\
\nonl \hspace*{10mm} A confirmed case's or an infected area's signal profile $W_2$;  \\
\nonl \hspace*{10mm} Virus lifespan $\{\tau_i|i = 1,2,...,|W_2| - 1\}$;  \\
\nonl \hspace*{10mm} A proximity threshold $\alpha$.\\
\textbf{Output}: results of contact detection at different timestamps.
\BlankLine
Initialize $S$ to empty; \\
Construct the processed profile  $\hat{W}$ from $W_2$ and $\{\tau_i|i = 1,2,...,|W_2| - 1\}$;\\

\ForEach{($A_i, t_i) \in W_1$}
{
    contact = False;\\
    \ForEach{$(\hat{A}_j,t_j^{start}, t_j^{end}) \in \hat{W}$}
    {
        \If{$t_j^{start} \leq t_i \leq t_j^{end}$} 
        {
            $s = P(A_i, \hat{A}_j)$;\\
            \If{$s \geq \alpha$}{ contact = True;\\
            break;}
        }
    }
    \uIf{contact == True}{Add $(True, t_i)$ to $S$;}
    \uElse{Add $(False, t_i)$ to $S$;}
    
}

\textbf{return} $S$

\caption{Contact Detection}
\label{algo_indirect_tracing}
\end{algorithm2e}

%% file: 6_evalution.tex
\section{Illustrative Experimental Results}
\label{sec:evaluation}
We have implemented and packaged \sysname{} as a Software Development Kit (SDK). In this section, we present illustrative experimental results on the SDK, 
using phones as IoT devices.
We first introduce the experiment settings in Section~\ref{sec:settings}. Then we study
how to set the threshold $\alpha$ in Section~\ref{sec:threshold}. We present 
the performance of \sysname{} for patients with app and infected areas in different sites in Sections \ref{sec:different_sites} and \ref{sec:in-out-detection}, respectively. Then, we compare \sysname{} with other state-of-the-art approaches in Section~\ref{sec:comparison_with_ble}. The studies on the impacts of different AP numbers, dynamic environment and heterogeneous devices are covered in Sections~\ref{sec:number_of_signals},~\ref{sec:dynamic environment} and~\ref{sec:heterogeneous_devices}, respectively. Finally, we discuss the impact of data sampling rate in Section \ref{sec: data_sampling_rate}.

\subsection{Experimental settings}
\label{sec:settings}
We
collect WiFi data using five mobile phones in three different sites. The brands and models of phones are different, and include Honor, Huawei Nova, Huawei Mate30, Xiaomi, and OPPO. According to some latest reports, these brands are representative in the market. The three experimental sites are an office, a bus station, and a store in a shopping mall. The size of the office is around $10$m$\times 12$m. The bus station is an outdoor area, the size of which is around $2$m$\times 15$m. The area in the shopping mall for experiments is a large store with a size of $20$m$\times 25$m. The total signal numbers are $32$ in the office, $109$ in the bus station, and $301$ in the shopping mall. The average number of signals~(i.e., scanned APs) in signal vectors of the office, bus station, and shopping mall are $19.02$, $24.0$, $46.29$, respectively.

To evaluate the detection performance for the case where the signal profiles of confirmed cases are available, we first put the five mobile devices at a location $\ell_0$ for $10$ minutes to collect the WiFi data in each site. The WiFi signals with RSSIs scanned by a device are collected. Then we put the devices at a location $\ell_i$ for $10$ minutes for data collection, where $i = 1,2,3,4,5,6,7,8,9,10$, and the distance between $\ell_0$ and $\ell_i$ is $i$ meters. The data sampling rate is set as $5$s per record, so we have around $120$ records of data for a device in each distance setting for each site.

\begin{figure*}
\noindent\makebox[\linewidth][c]{
    \subfigure[\small k = 1m.]{
        \label{fig:1m}
        \includegraphics[width=0.33\textwidth]{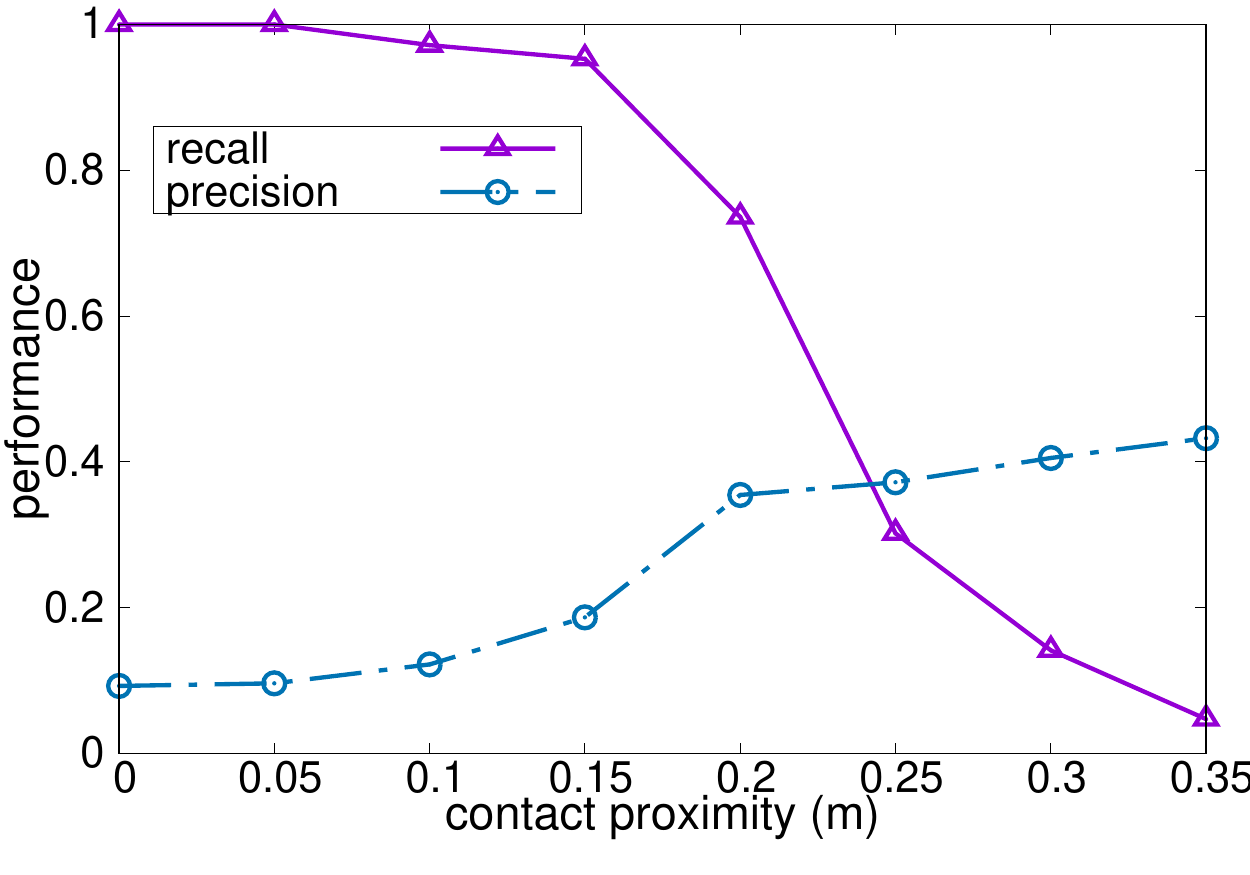}
    }
    
    \subfigure[\small k = 2m.]{
        \label{fig:2m}
        \includegraphics[width=0.33\textwidth]{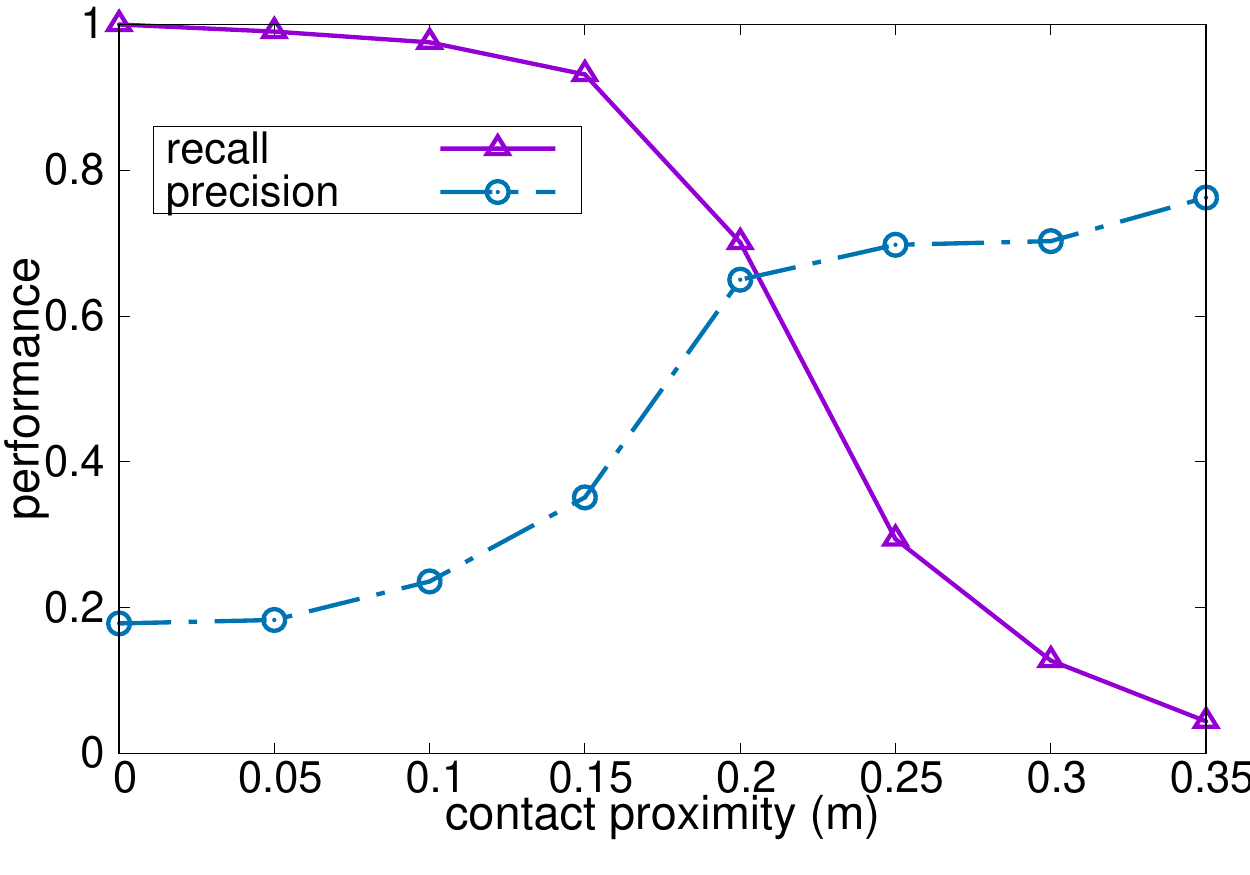}
    }
     \vspace{.6in}
    \subfigure[\small k = 4m.]{
        \label{fig:4m}
        \includegraphics[width=0.33\textwidth]{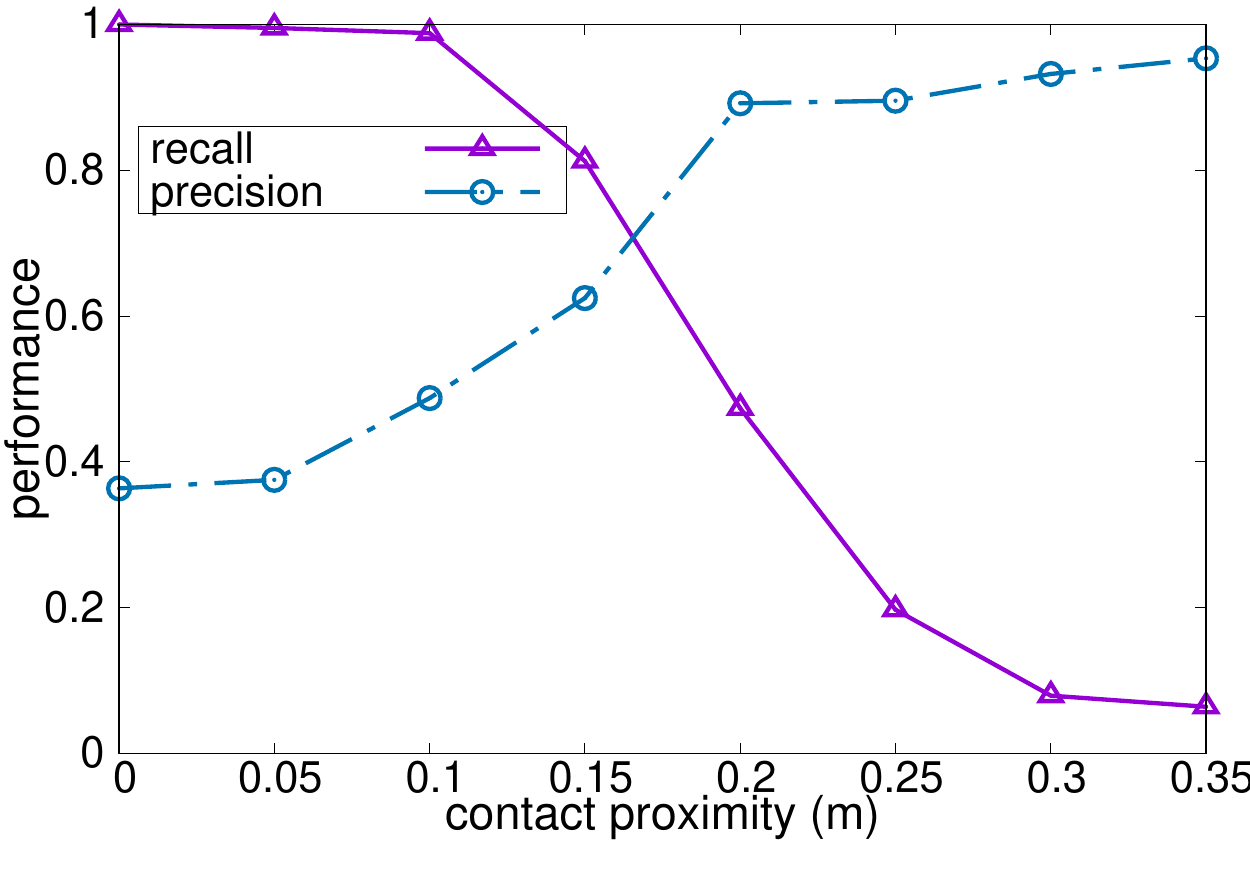}
    }
}
\caption{Precision and Recall for different contact proximity $K$.}
\label{fig:different_distance}
\end{figure*}

To evaluate the detection performance for the case where a confirmed case's signal profile is unavailable, we walk in the experimental sites to collect WiFi data using a mobile phone to construct the processed profiles for each site. Then we wander around and outside the area with five mobile phones collecting WiFi data for testing. The time when we were in and outside the area is recorded during the experiments.

Given the data $D$ collected by a user's device, we use $D_a$ to denote the data which are collected when the user has contact with the virus~(i.e., within the contact proximity with a confirmed case or in an infected area), and use $D_b$ to denote the data which are detected as having contact with the virus. The $D_a$ is the ground-truth data while the $D_b$ is the detection result. Precision, recall, and F1-score are used as metrics to evaluate the contact detection results. The precision is defined as 
\begin{equation}
    precision = \frac{|D_a \cap D_b|}{|D_b|},
\end{equation}
 where $|\cdot|$ represents the data size. Similarly, recall is defined as 
 \begin{equation}
     recall = \frac{|D_a \cap D_b|}{|D_a|}.
 \end{equation}
Based on the definition of precision and recall, F1-score is defined as 
\begin{equation}
    F_1 = 2 * \frac{precision * recall}{precision + recall}.
\end{equation}
 
We compare \sysname{} with some other state-of-the-art approaches, which are introduced as follows,
\begin{itemize}
    \item {\em Bluetooth:} It is widely used for digital contact tracing, such as schemes~\cite{MIT_PACT}\cite{chan2020pact}\cite{troncoso2020decentralized}. To collect Bluetooth data, two mobile devices are put at a distance of $k$ meters for $10$ minutes in the three experimental sites, where $k$ is set to be $\{1$, $2$, \dots, $10\}$. We use one device as the broadcaster, and another as the scanner. The scanner can scan the Bluetooth signal from the broadcaster, and 
   the RSSI is recorded over time. For each contact proximity $k$ meters, a threshold is selected for contact detection. If a received signal strength is larger than the threshold, they are detected as having contact. 
    \item {\em Jaccard similarity:} It is used to evaluate the similarity of two sets, and it is defined as the size of the intersection divided by the size of the union of two sets. If the Jaccard similarity of two signal vectors is larger than a threshold, they are identified as within the contact proximity. It is also used in a relevant work for proximity estimation~\cite{dmitrienko2020proximity}. 
    \item {\em Average Manhattan distance~(AMD):} It is used in previous works~\cite{sapiezynski2017inferring}~\cite{dmitrienko2020proximity}, which is defined as
    \begin{equation}
        AMD = \frac{\sum_i |RSSI_{A,i} - RSSI_{B,i}|}{N},
    \end{equation}
    where $RSSI_{A,i}$ is the received signal strength of AP $i$ measured by user A, and N is the total number of overlapping APs.
    If the AMD of two signal vectors is less than a threshold, they are identified as within the contact proximity. 
    \item {\em Average Euclidean distance~(AED):} It is also used in the previous work~\cite{sapiezynski2017inferring}, which is defined as 
    \begin{equation}
        AED = \frac{\sqrt{\sum_i (RSSI_{A,i} - RSSI_{B,i})^2}}{N},
    \end{equation}
    where $RSSI_{A,i}$ is the received signal strength of AP $i$ measured by user A, and N is the total number of overlapping APs.
    If the AED of two signal vectors is less than a threshold, they are identified as within the contact proximity.
\end{itemize}

For the baseline approaches AMD and AED, given two signal vectors $A$ and $B$, if a signal is scanned in $A$ but not in $B$, the signal strength is set as -$100$ in $B$ for calculation, and vice versa.
 
\subsection{Threshold $\alpha$}
\label{sec:threshold}

As mentioned in Section~\ref{sec:contact_detection}, the contact detection algorithm relies on a threshold $\alpha$ to identify contacts. In this section, we discuss the selection of $\alpha$. Given the contact proximity $k$m, if the distance of a user and the virus is less than $k$m, she/he should be detected as having contact with the virus. Intuitively, $\alpha$ is relevant to the contact proximity and it should be different for different contact proximities. We use the data collected at $\ell_0$ in a site as the data from confirmed cases, and detect contacts for data which are collected at $\ell_i$~($i > 0$) in the same site. When $k$ meters is set as the contact proximity, $D_a$ contains the data collected at $\ell_i$ where $i \leq k$. 

\begin{figure*}
\noindent\makebox[\linewidth][c]{
    \subfigure[\small Precision.]{
        \label{fig:precision_different_site}
        \includegraphics[width=0.33\textwidth]{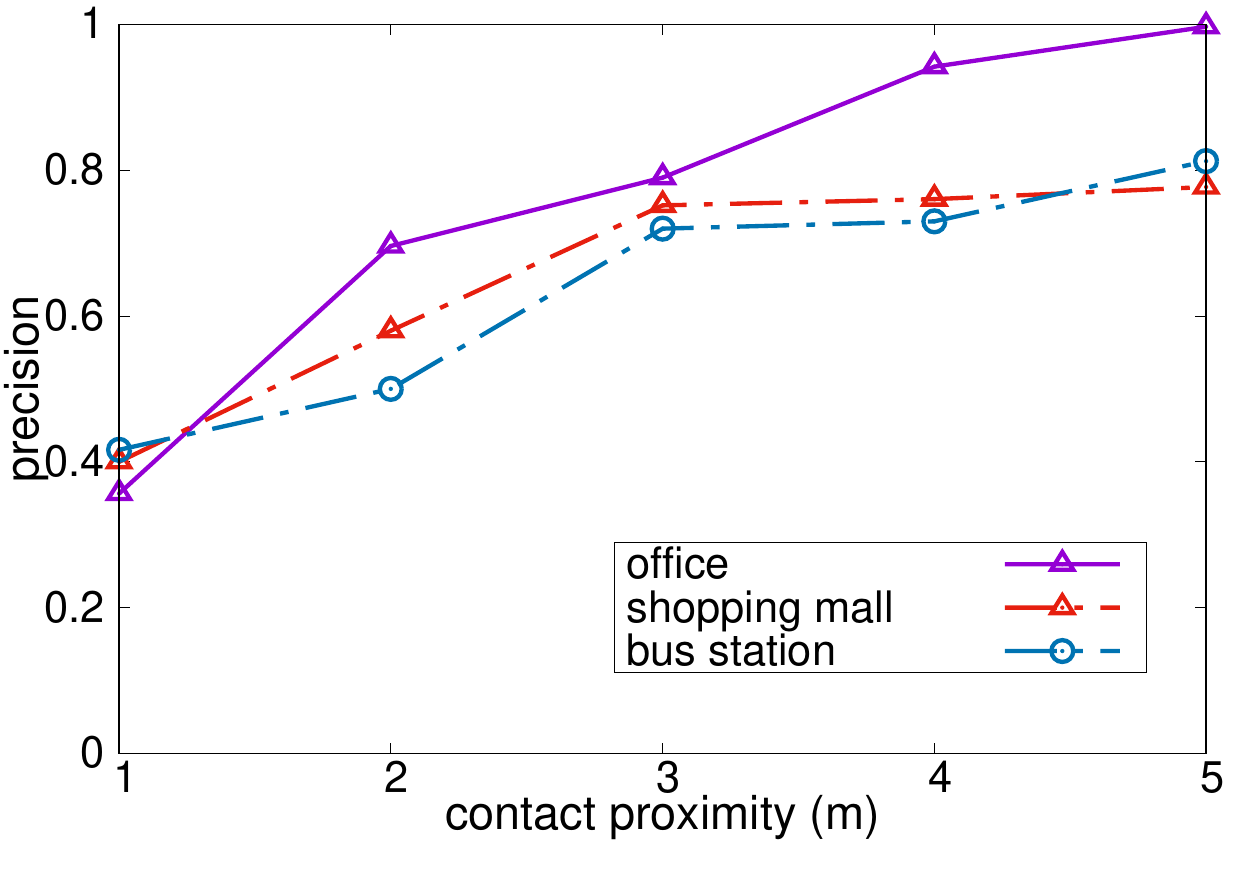}
    }
    \subfigure[\small Recall.]{
        \label{fig:recall_different_site}
        \includegraphics[width=0.33\textwidth]{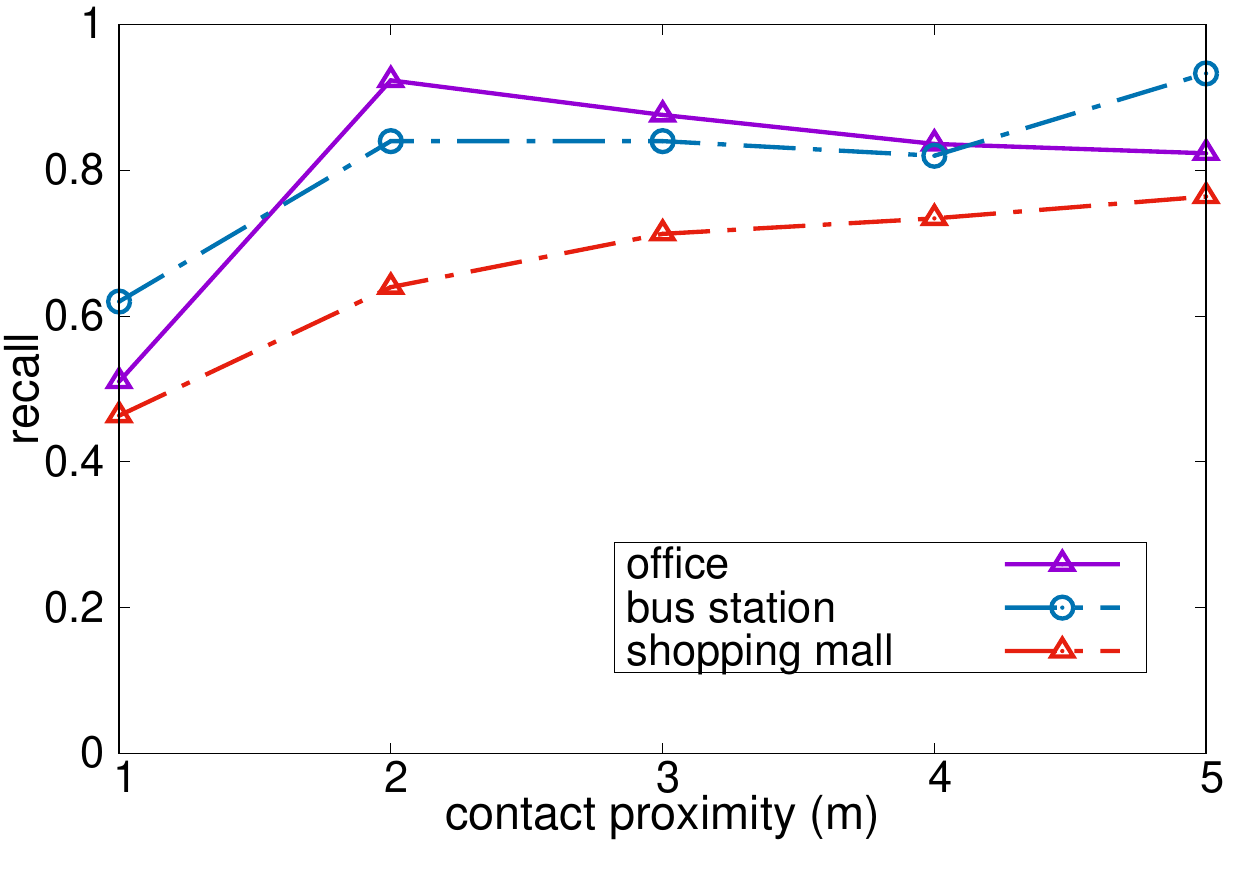}
    }
    \subfigure[\small F1-score.]{
        \label{fig:f1_different_site}
        \includegraphics[width=0.33\textwidth]{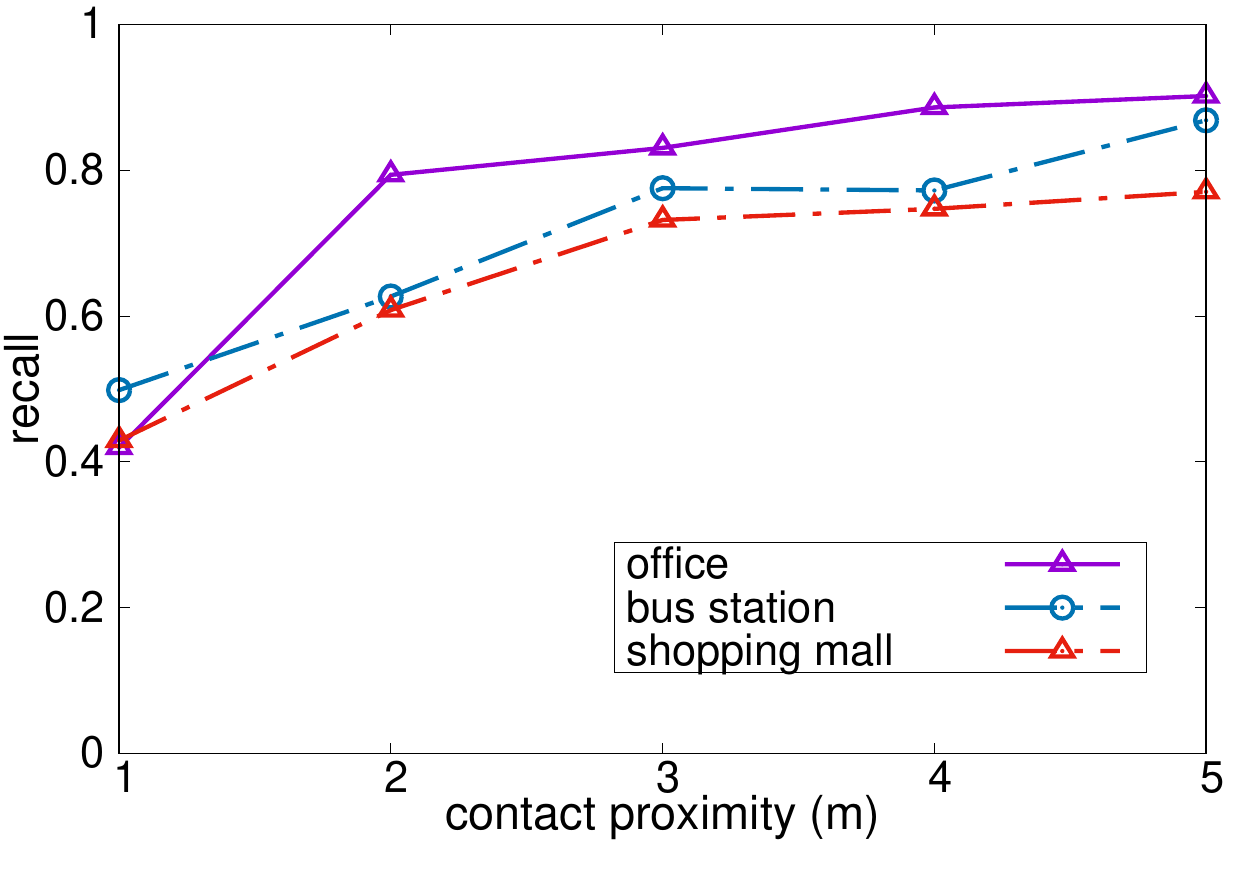}
    }
}
\caption{Performance in different sites.}
\label{fig:performance_in_different_sites}
\end{figure*}

Precision and recall are used as metrics, and the results of $\alpha$ versus precision and recall for $k = 1$m, $k = 2$m and $k = 4$m are presented in Figure \ref{fig:different_distance}. 
As the threshold $\alpha$ increases, the precision increases while the recall declines. The reason is that a larger threshold indicates closer proximity. Thus, increasing the threshold would lead to high precision. However, if the threshold is set to be too large, some of the data the distance of which is less than $k$m will not be detected, resulting in a drop in recall. 

The threshold can be selected according to the requirements of precision and recall for close contact detection. To balance the precision and recall, we select the intersection points, the precision and recall of which are equal for our following discussion. 
In Figure \ref{fig:1m}, the precision and recall for $k = 1$m are low when $\alpha$ is set as $0.25$, which indicates that identifying contact within $1$m is difficult.  As shown in Figures \ref{fig:2m}, the precision and recall for $k = 2$m have a significant improvement when the threshold is around $0.20$.  The precision and recall in Figure \ref{fig:4m} for $k = 4$m are high ~(around $70\%$) if the threshold is around $0.17$. We use the same strategy to select thresholds for other contact proximities.

\subsection{Site study}
\label{sec:different_sites}

We present the performance of contact detection in different sites in this section. We use different distances~($k = 1$m, $2$m, $3$m, $4$m, $5$m) to denote the contact proximity, and the threshold is set according to the discussion in Section \ref{sec:threshold}. Results of precision, recall and F1-score versus contact proximity are shown in Figure \ref{fig:performance_in_different_sites}.

In Figure \ref{fig:precision_different_site}, as the contact proximity increases, the precision in the three sites increases, indicating that it is easier to detect contacts within a greater proximity. The precision for $k = 1$m is low in all sites. The result shows the difficulties of identifying whether the contact happens in $1$m because the WiFi signals within a $1$-m range are usually similar. However, the precision has significant improvements for larger contact proximity. 
The precision is high~($50\%$ -- $70\%$) when the proximity is $2$m. 
The precision indoors~(office and shopping mall) is better than the precision outdoors because WiFi signals are more stable indoors. The improvement is more significant in the office scenario compared with the shopping mall scenario. The recall shown in \ref{fig:recall_different_site} indicates the good performance of \sysname{} to detect those who have close contact.  We present the F1-score result in Figure \ref{fig:f1_different_site}, indicating the satisfactory overall performance of \sysname{}. 

\subsection{In-out detection of infected areas}
\label{sec:in-out-detection}

\begin{figure}
    \centering
  \includegraphics[width=.65\linewidth]{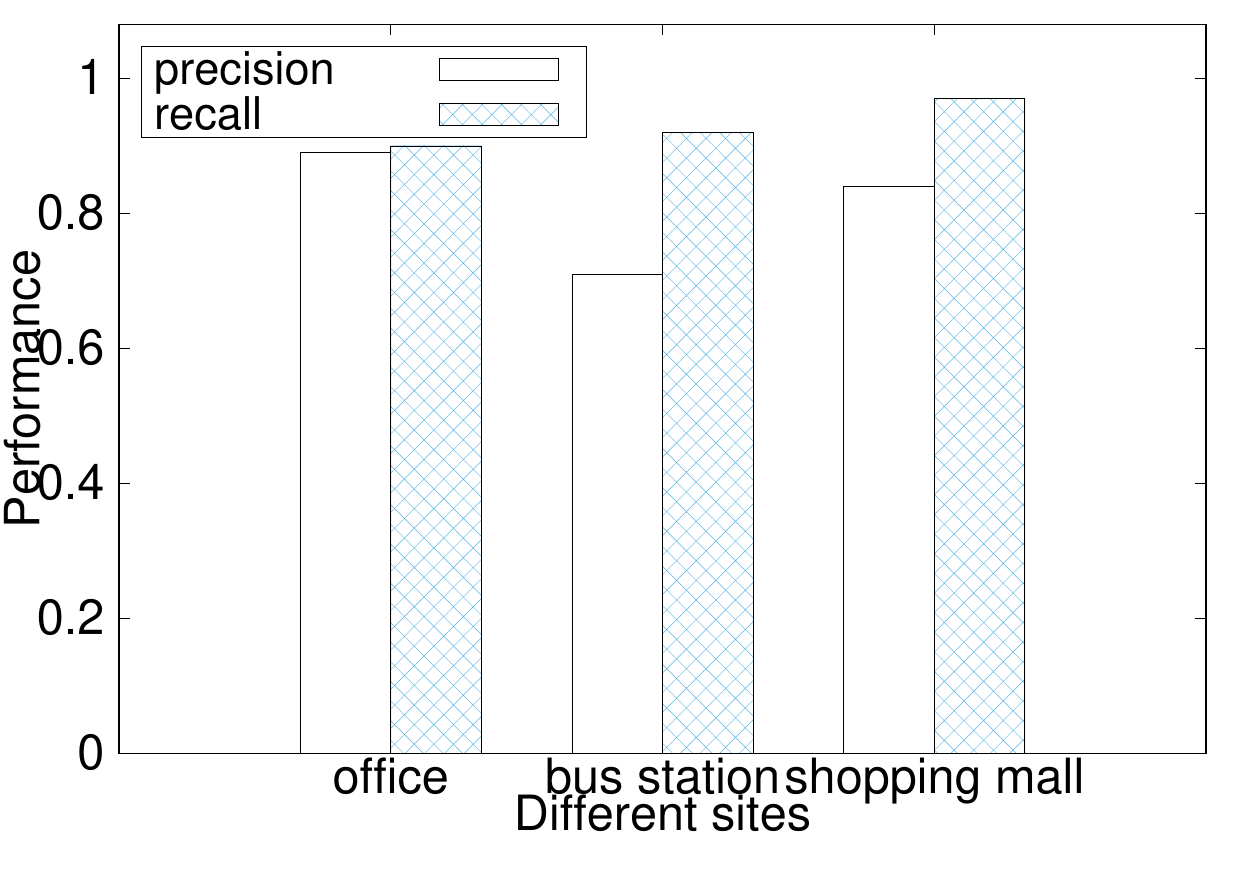}
  \caption{Precision and recall of in-out detection. }
  \label{fig:area_detection}
\end{figure}

Contact detection for confirmed cases without the app is to detect whether a user has been in or outside an infected area. We construct processed profiles for the office, bus station, and a store in a shopping mall using the collected WiFi data. Then we compare the similarity between the processed profile of the area and the data collected in and outside the area. If the similarity is larger than the  threshold $\alpha$, the data are identified as being collected in the area and having contact with the virus. $\alpha$ is set as $0.2$ in the experiment. Precision and recall are used as the metrics for evaluation.  The results are shown in Figure \ref{fig:area_detection}. The detection in all the sites achieves good performances. The precision and recall are high for the three sites, illustrating that \sysname{} is very efficient for in-out detection of infected areas.


\subsection{Comparison with other approaches}
\label{sec:comparison_with_ble}

\begin{figure*}
\noindent\makebox[\linewidth][c]{
    \subfigure[\small Precision.]{
        \label{fig:comparison_office_precision}
        \includegraphics[width=0.33\textwidth]{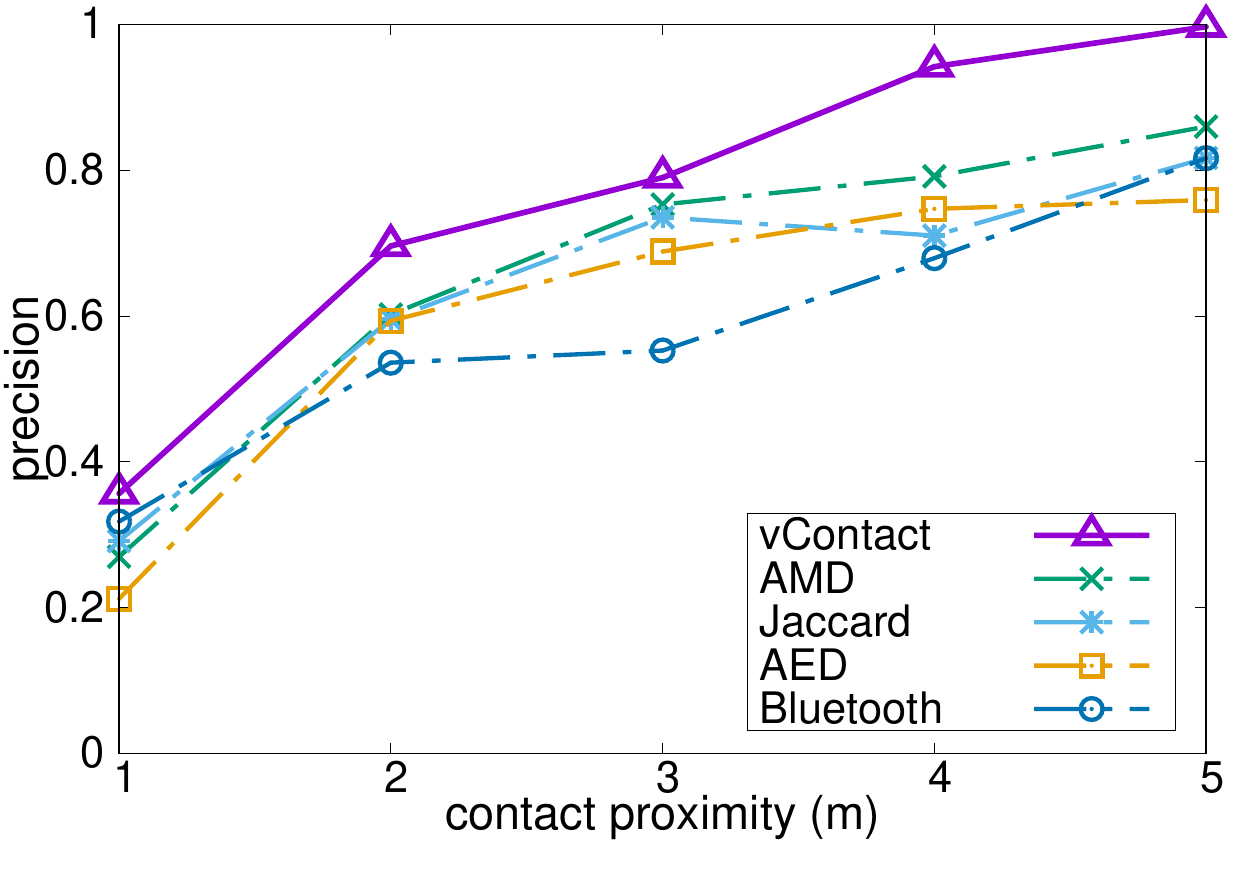}
    }
    
    \subfigure[\small Recall.]{
        \label{fig:comparison_office_recall}
        \includegraphics[width=0.33\textwidth]{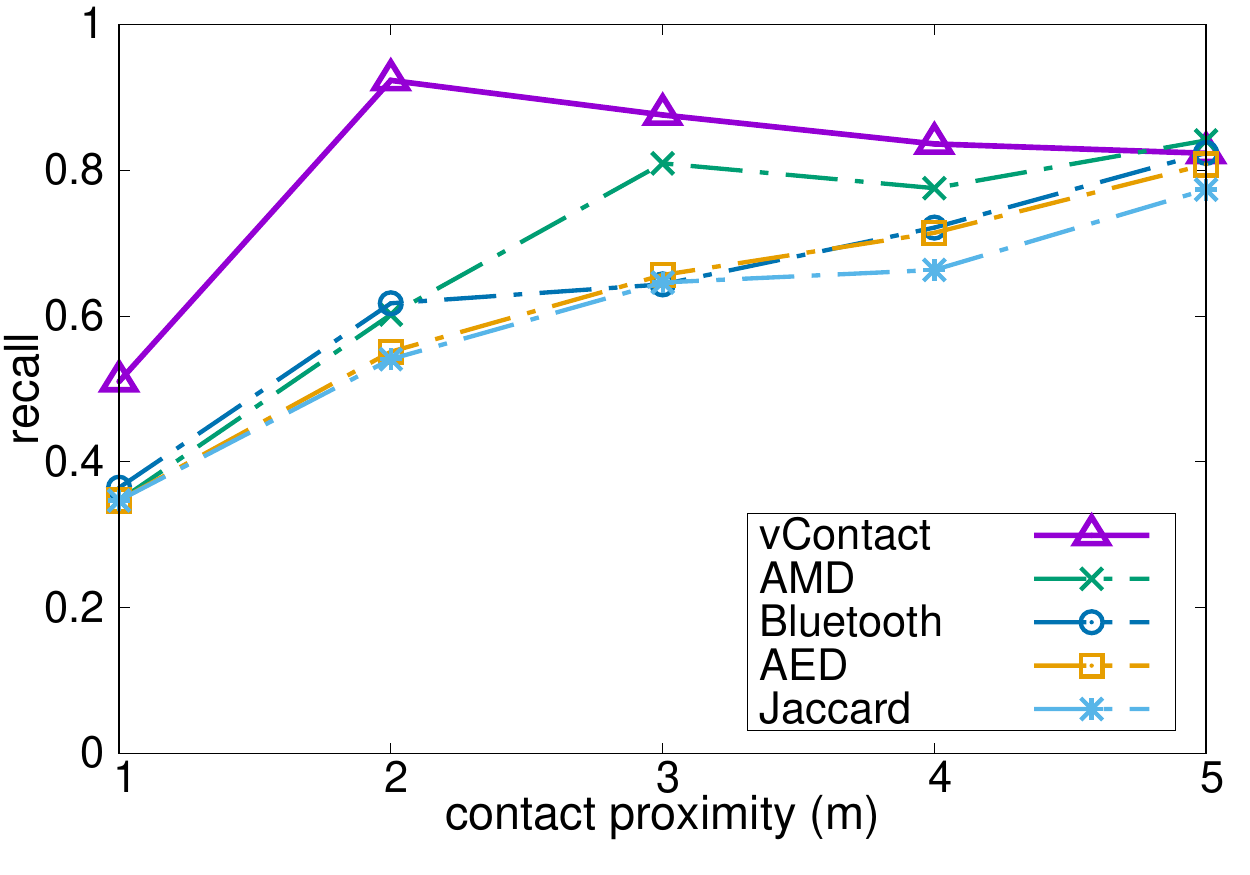}
    }
     \vspace{.6in}
    \subfigure[\small F-1 score.]{
        \label{fig:comparison_office_F1}
        \includegraphics[width=0.33\textwidth]{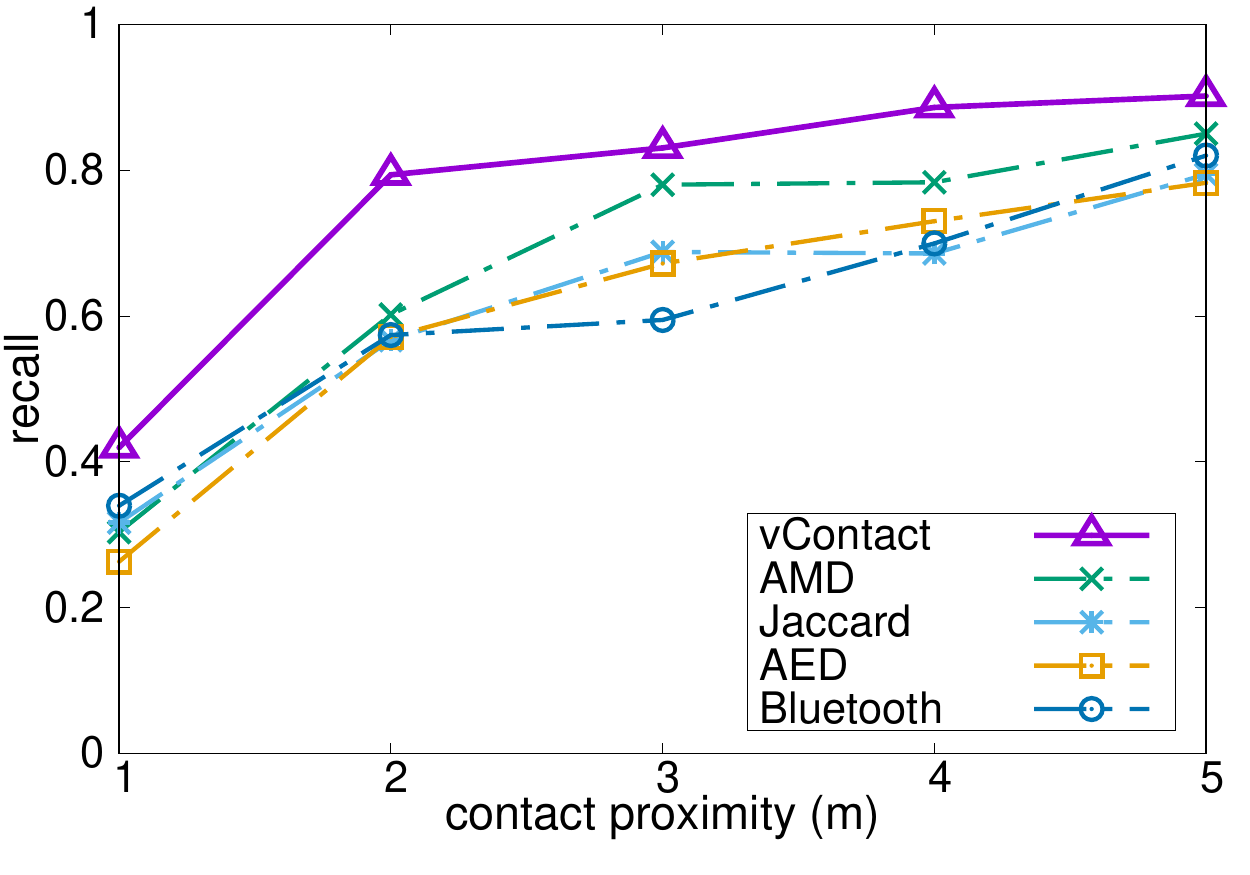}
    }
}
\caption{Comparison with baseline approaches on the office dataset.}
\label{fig:comparison_baseline_office}
\end{figure*}

\begin{figure*}
\noindent\makebox[\linewidth][c]{
    \subfigure[\small Precision.]{
        \label{fig:comparison_bus_station_precision}
        \includegraphics[width=0.33\textwidth]{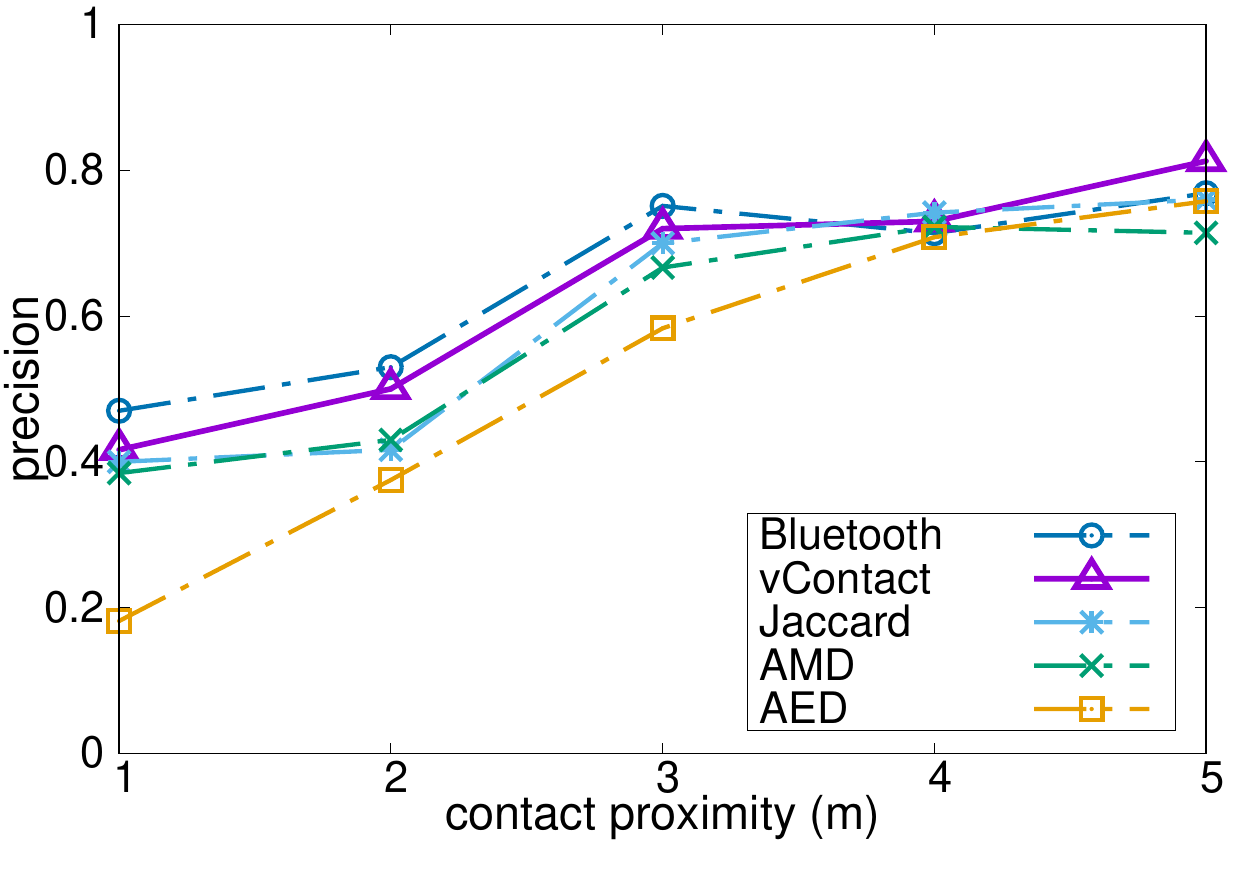}
    }
    
    \subfigure[\small Recall.]{
        \label{fig:comparison_bus_station_recall}
        \includegraphics[width=0.33\textwidth]{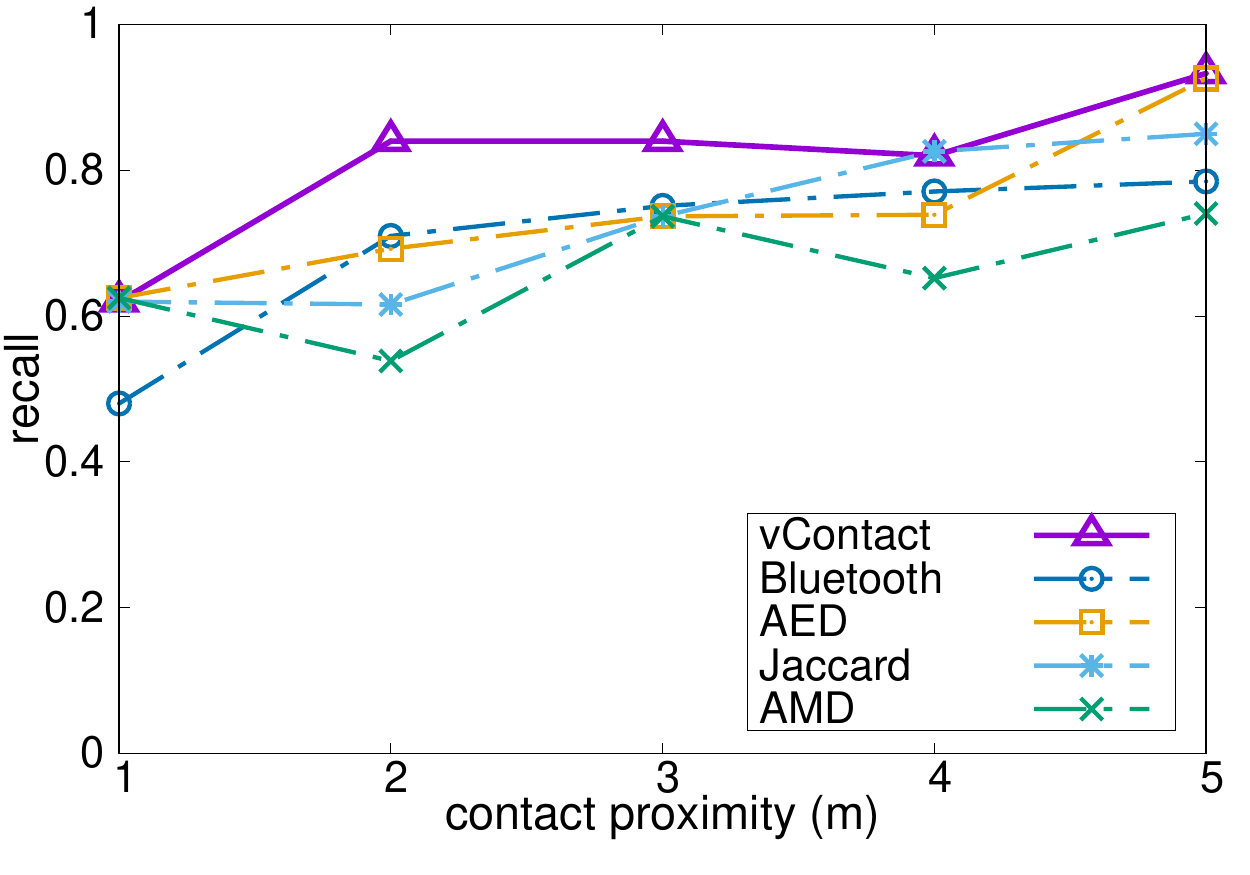}
    }
     \vspace{.6in}
    \subfigure[\small F-1 score.]{
        \label{fig:comparison_bus_station_F1}
        \includegraphics[width=0.33\textwidth]{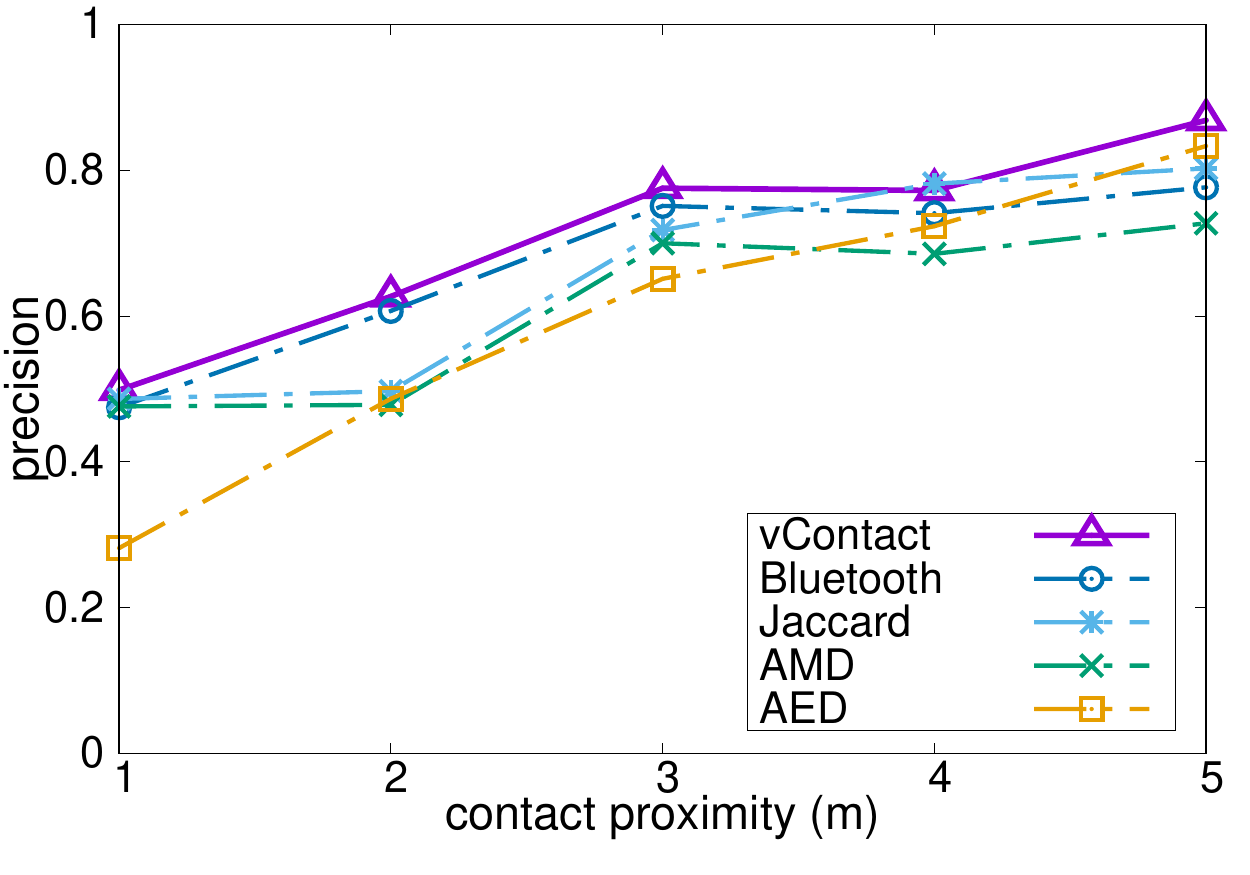}
    }
}
\caption{Comparison with baseline approaches on the bus station dataset.}
\label{fig:comparison_baseline_bus_station}
\end{figure*}

\begin{figure*}
\noindent\makebox[\linewidth][c]{
    \subfigure[\small Precision.]{
        \label{fig:comparison_shopping_mall_precision}
        \includegraphics[width=0.33\textwidth]{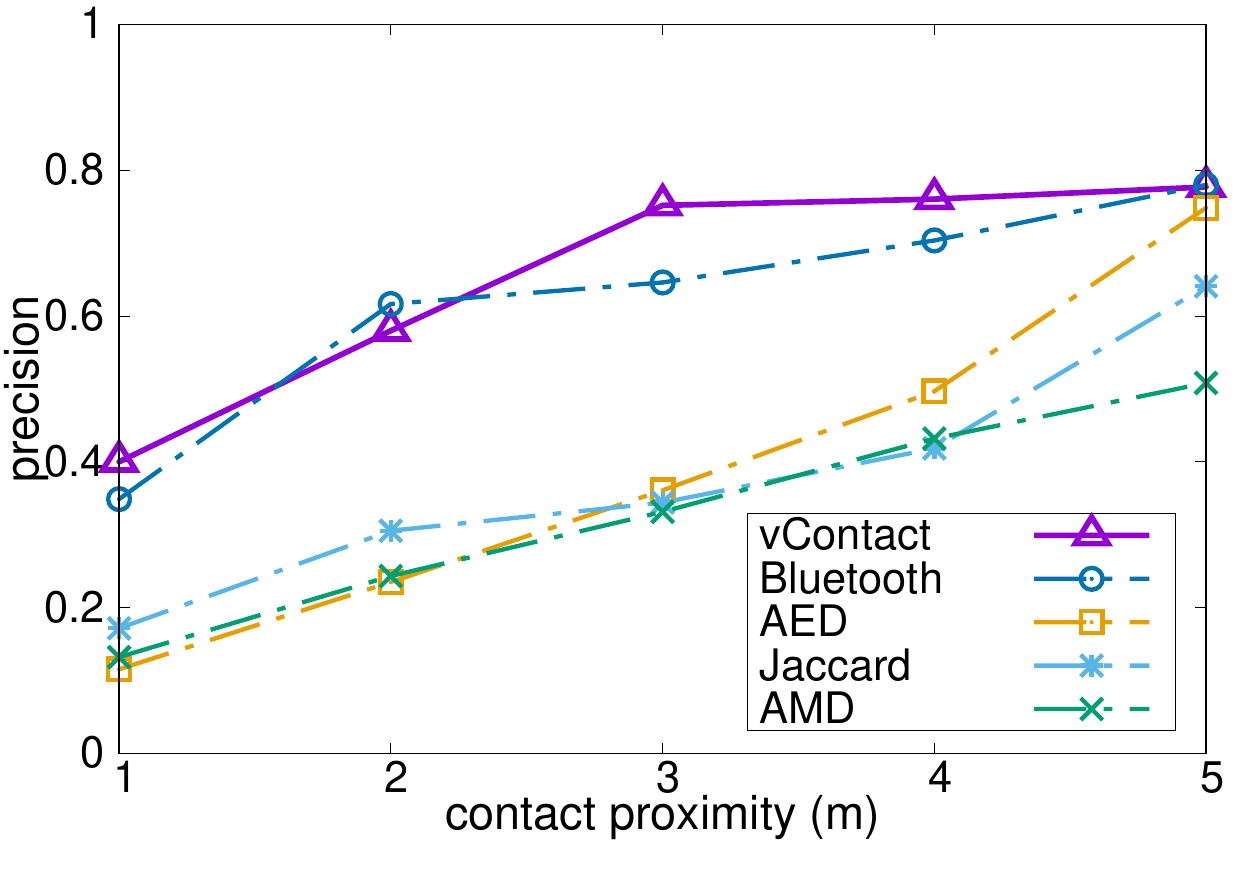}
    }
    
    \subfigure[\small Recall.]{
        \label{fig:comparison_shopping_mall_recall}
        \includegraphics[width=0.33\textwidth]{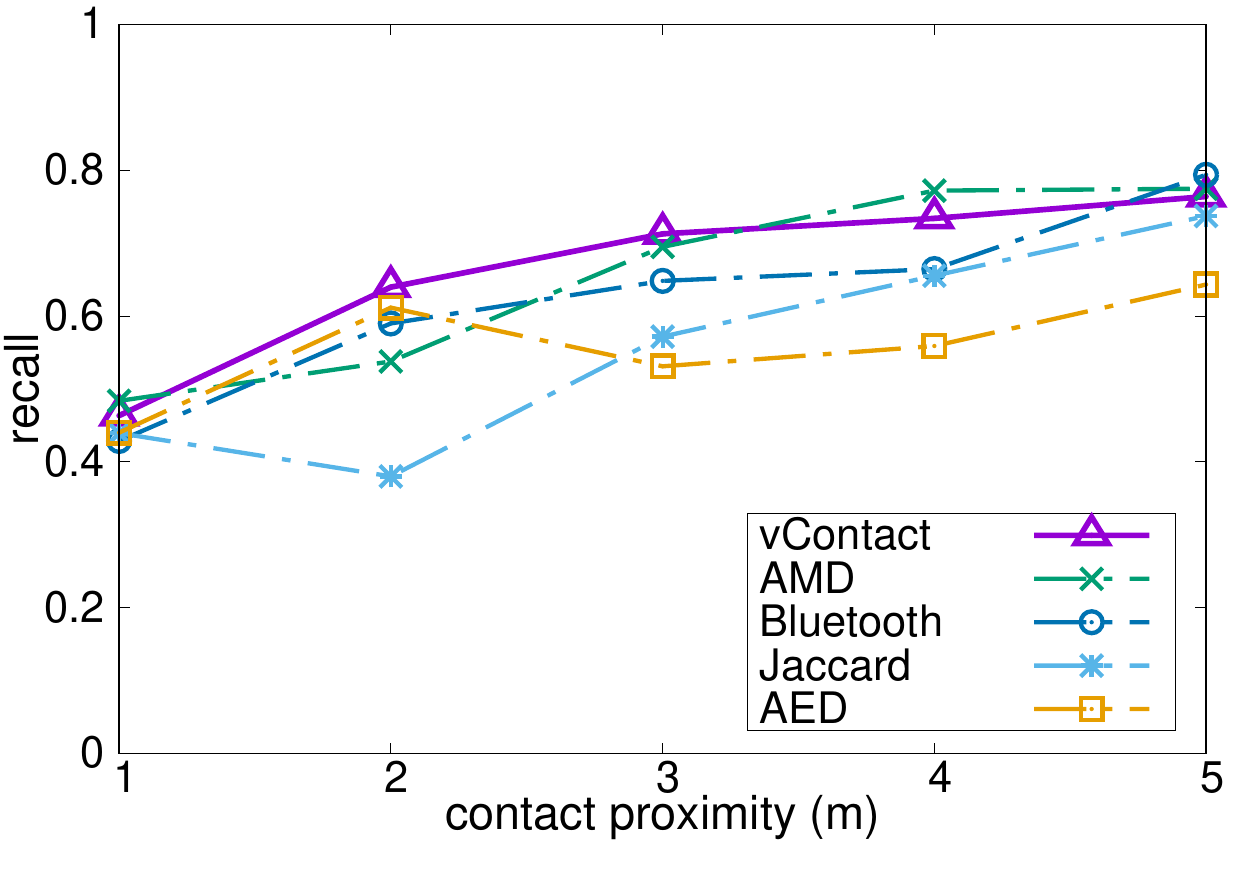}
    }
     \vspace{.6in}
    \subfigure[\small F-1 score.]{
        \label{fig:comparison_shopping_mall_F1}
        \includegraphics[width=0.33\textwidth]{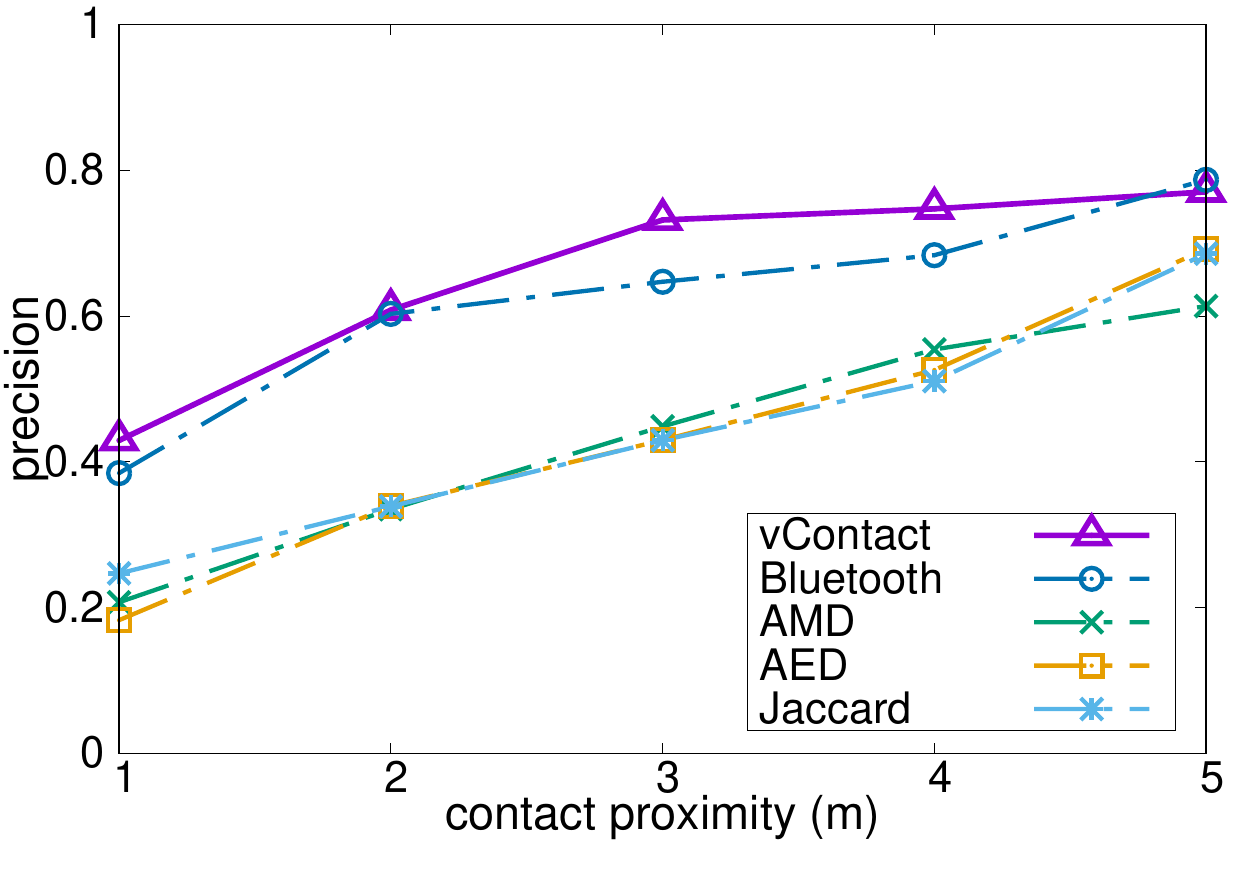}
    }
}
\caption{Comparison with baseline approaches on the shopping mall dataset.}
\label{fig:comparison_baseline_shopping_mall}
\end{figure*}

 As the baseline approaches rely on a selected threshold to detect contact, for a given contact proximity, we use the same strategy to select thresholds as discussed in Section \ref{sec:threshold}. Precision, recall, and F1-score are used as metrics for performance comparison.

The results of  precision, recall and F1-score  versus proximity on the three datasets are presented in Figures \ref{fig:comparison_baseline_office}~(the office), \ref{fig:comparison_baseline_bus_station}~(the bus station), and \ref{fig:comparison_baseline_shopping_mall}~(the shopping mall). In Figure \ref{fig:comparison_baseline_office}, the precision, recall, and F-1 score of different approaches increase as the contact proximity increases. \sysname{} always outperforms other baseline approaches on the metrics of precision and F-1 score. \sysname{} has higher recall than others when contact proximity is less than $5$m and has similar performance to Bluetooth when the contact proximity is $5$m. The curves of precision, recall and F1-score on the other datasets have a similar trend to that on the office dataset. As shown in Figure \ref{fig:comparison_bus_station_precision},
the precision of Bluetooth is slightly higher than \sysname{} on the bus station dataset. But \sysname{} has better performance than Bluetooth and other approaches with respect to recall and F1-score. As for the performance on the shopping mall dataset, \sysname{} has similar precision to Bluetooth when contact proximity is $1$m and $2$m, but has a significant improvement on precision when contact proximity is $3$m and $4$m. In Figure \ref{fig:comparison_shopping_mall_recall}, \sysname{} has similar recall to Bluetooth and AMD. \sysname{} always outperforms other approaches which use WiFi data for detection. 
Overall, \sysname{} has a higher F-1 score than other approaches in all datasets, indicating that it is more efficient for contact detection. We can also learn from the figures that \sysname{} and other approaches have better performance in the indoor scenario, and the improvement of \sysname{} is more significant compared with the outdoor site.

\subsection{AP number}
\label{sec:number_of_signals}
\begin{figure*}
\noindent\makebox[\linewidth][c]{
    \subfigure[\small Office.]{
        \label{fig:ap_num_indoor}
        \includegraphics[width=0.33\textwidth]{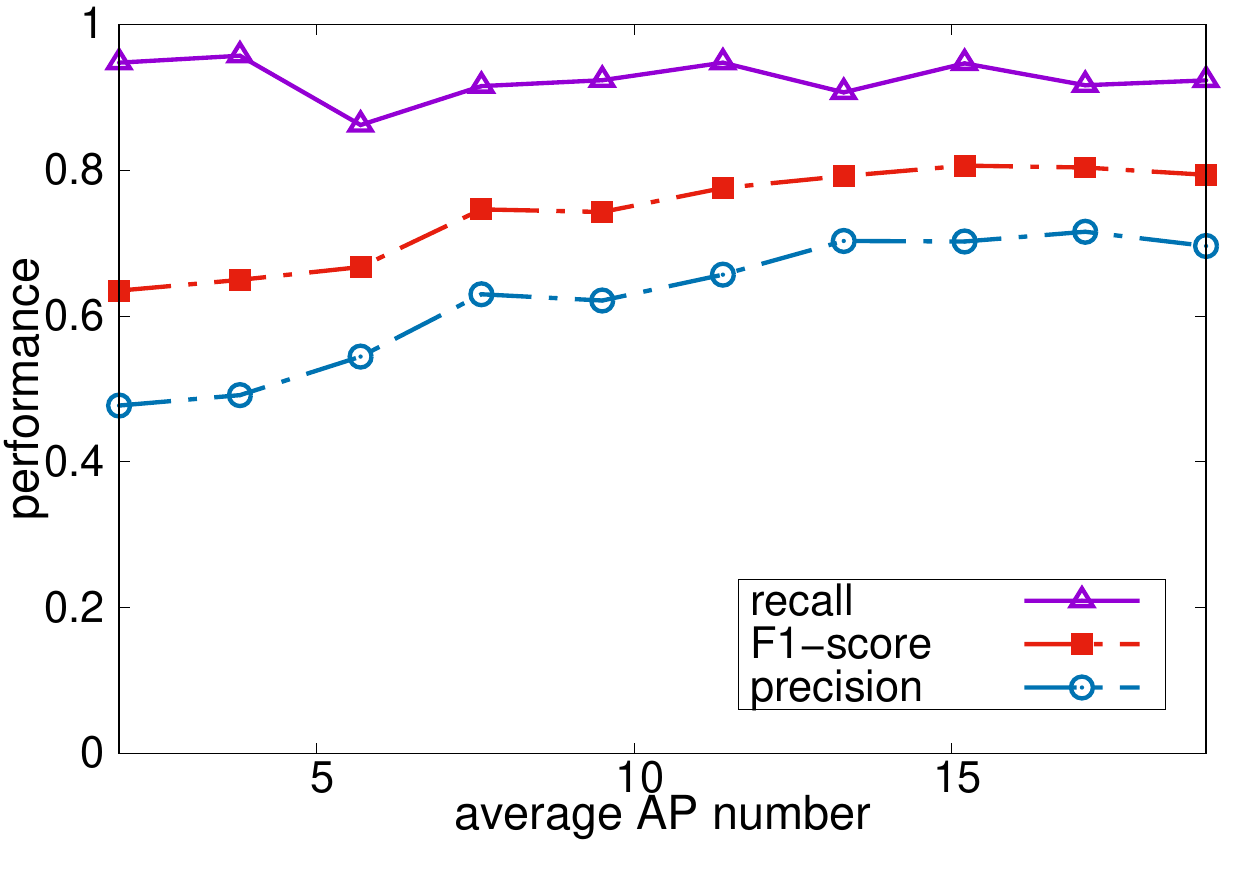}
    }
    
    \subfigure[\small Bus station.]{
        \label{fig:ap_mum_outdoor}
        \includegraphics[width=0.33\textwidth]{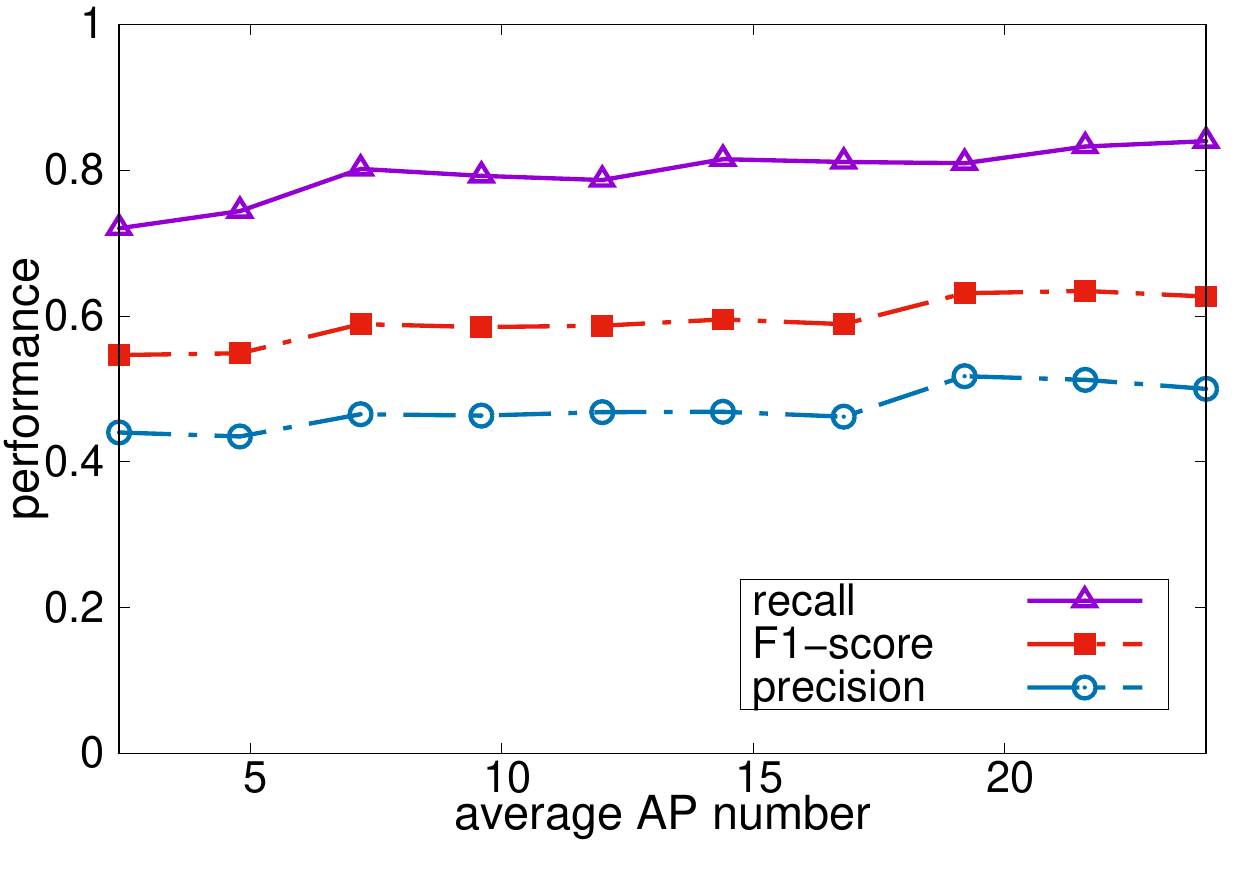}
    }
     \vspace{.6in}
    \subfigure[\small Shopping mall.]{
        \label{fig:ap_num_mall}
        \includegraphics[width=0.33\textwidth]{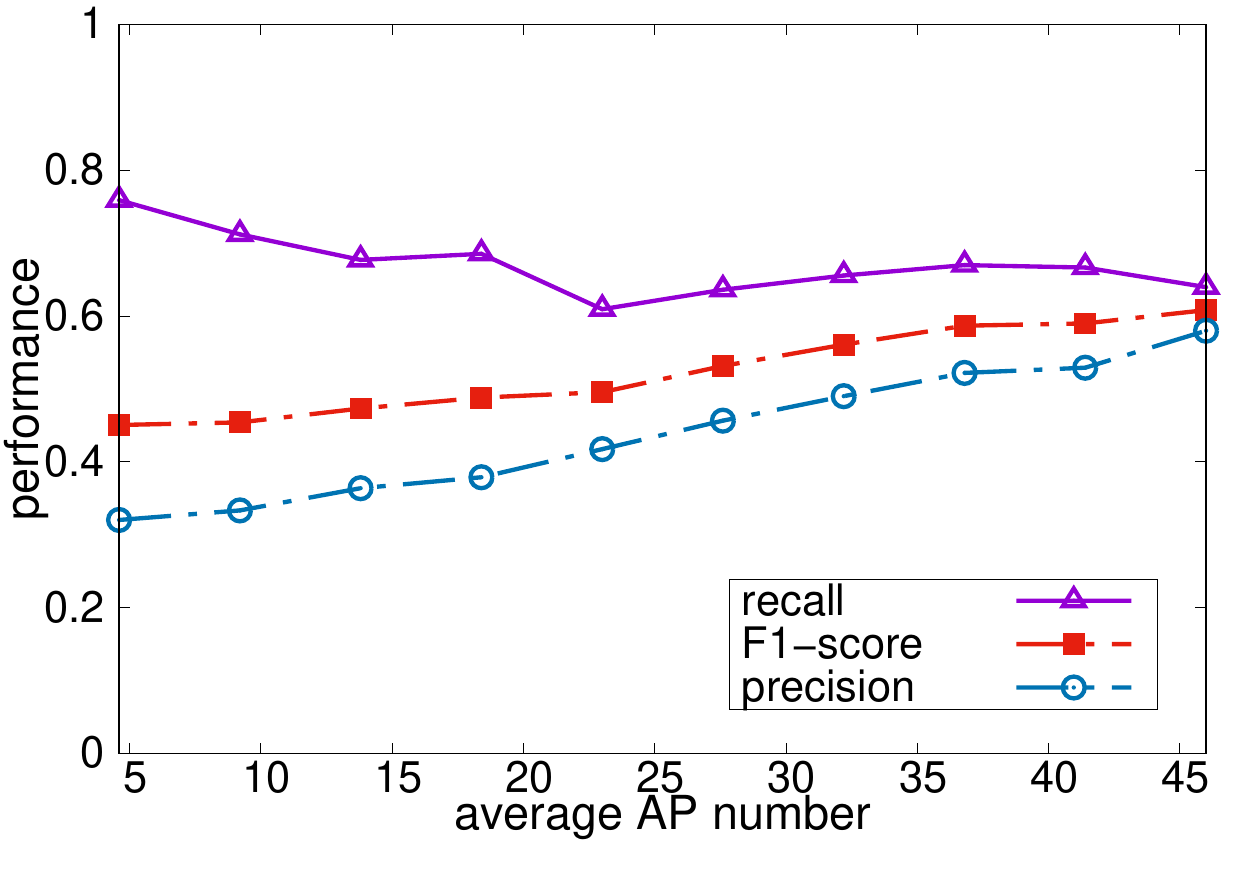}
    }
}
\caption{Impact of signal numbers (AP numbers) on the performance of contact detection.}
\label{fig:impact_ap_num}
\end{figure*}

\begin{figure*}
\noindent\makebox[\linewidth][c]{
    \subfigure[\small Precision.]{
        \label{fig:filtering_rate_precision}
        \includegraphics[width=0.33\textwidth]{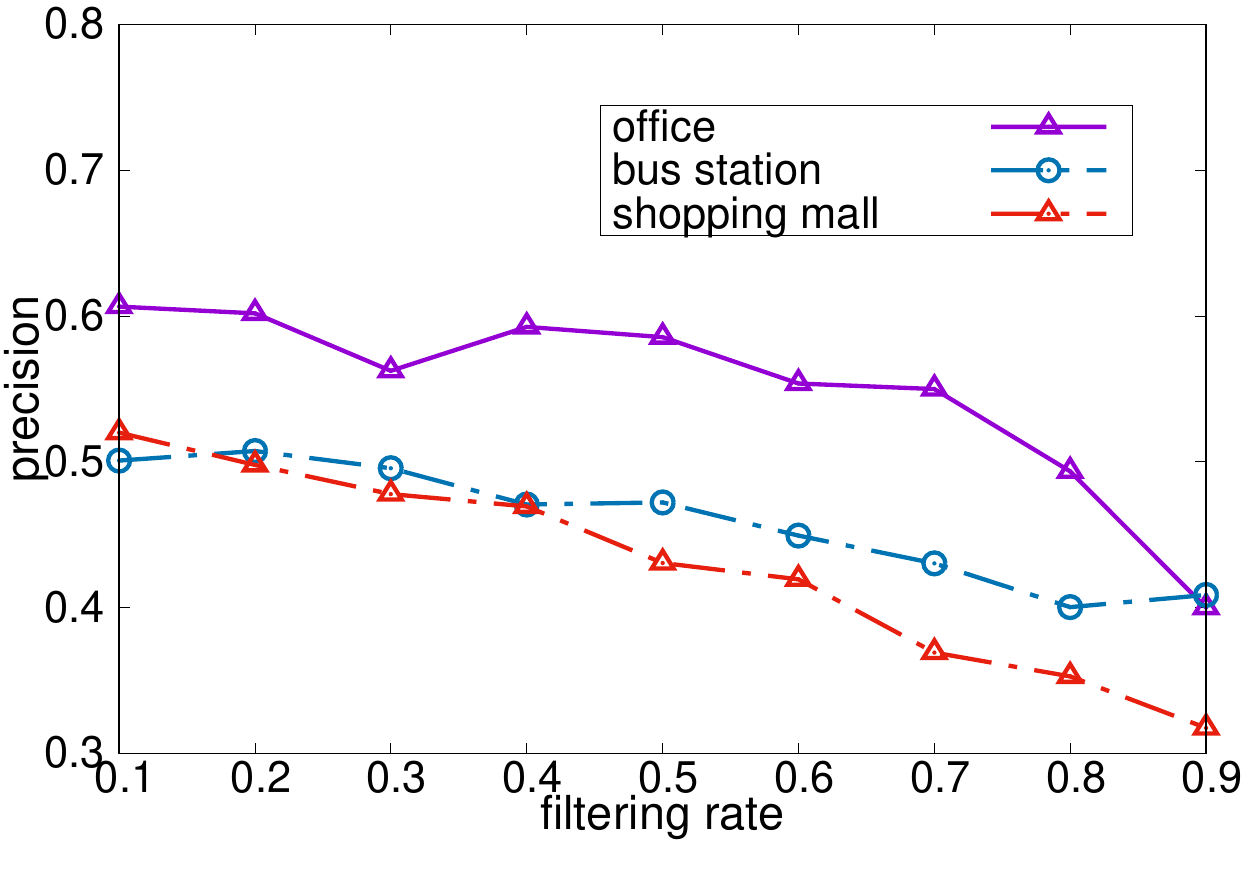}
    }
    
    \subfigure[\small Recall.]{
        \label{fig:filtering_rate_recall}
        \includegraphics[width=0.33\textwidth]{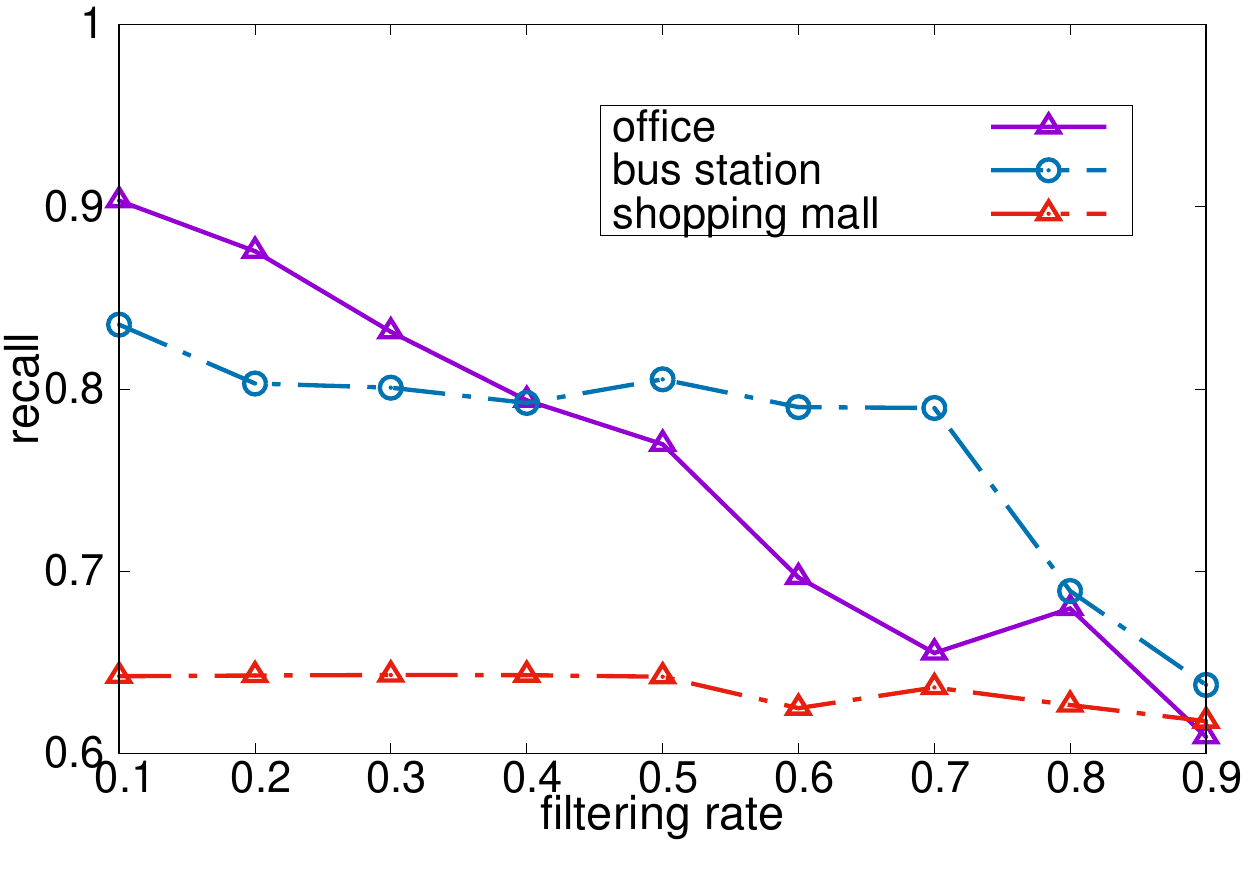}
    }
     \vspace{.6in}
    \subfigure[\small F1-score.]{
        \label{fig:filtering_rate_f1}
        \includegraphics[width=0.33\textwidth]{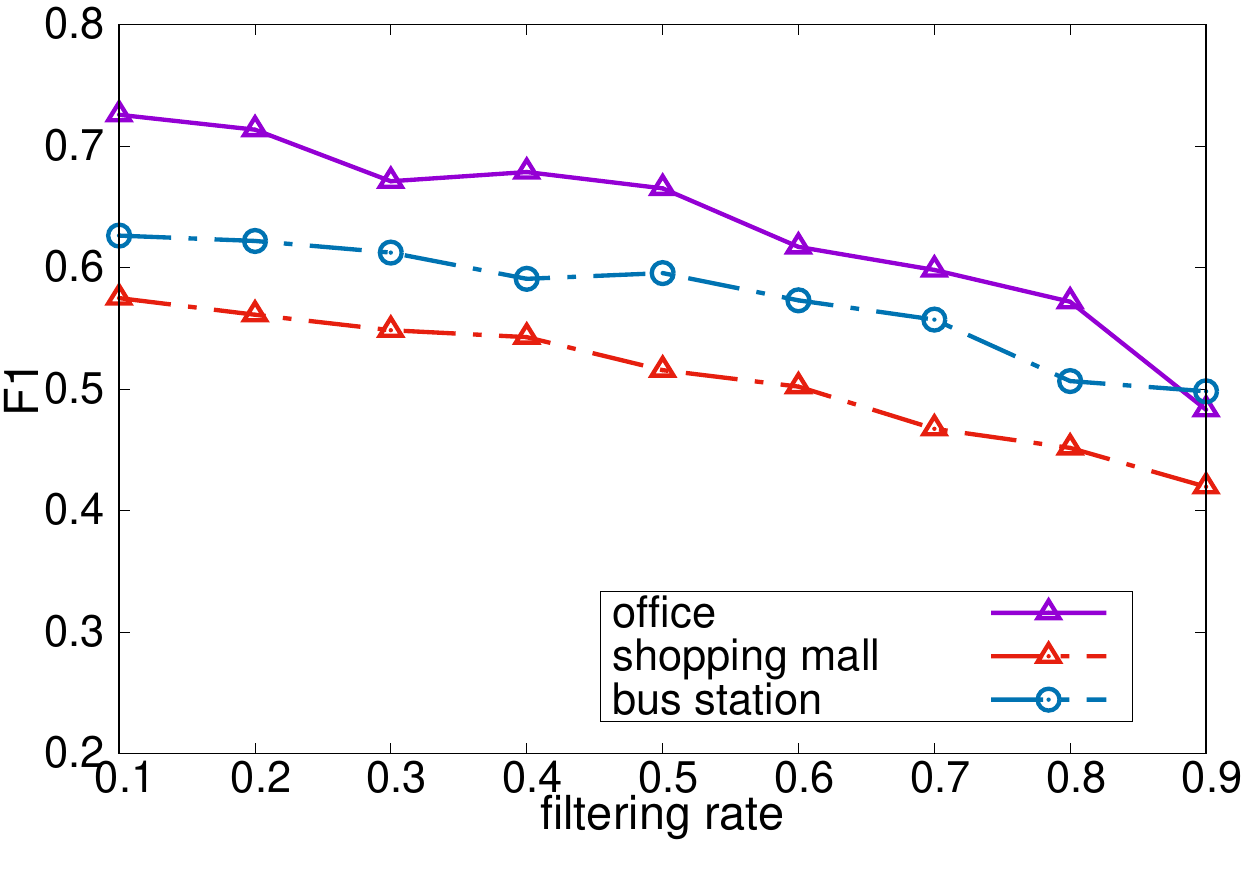}
    }
}
\caption{Impact of different filtering rates on the performance of contact detection.}
\label{fig:impact_filtering_rate}
\end{figure*}

\begin{figure*}
\noindent\makebox[\linewidth][c]{
    \subfigure[\small Precision.]{
        \label{fig:noise_precision}
        \includegraphics[width=0.33\textwidth]{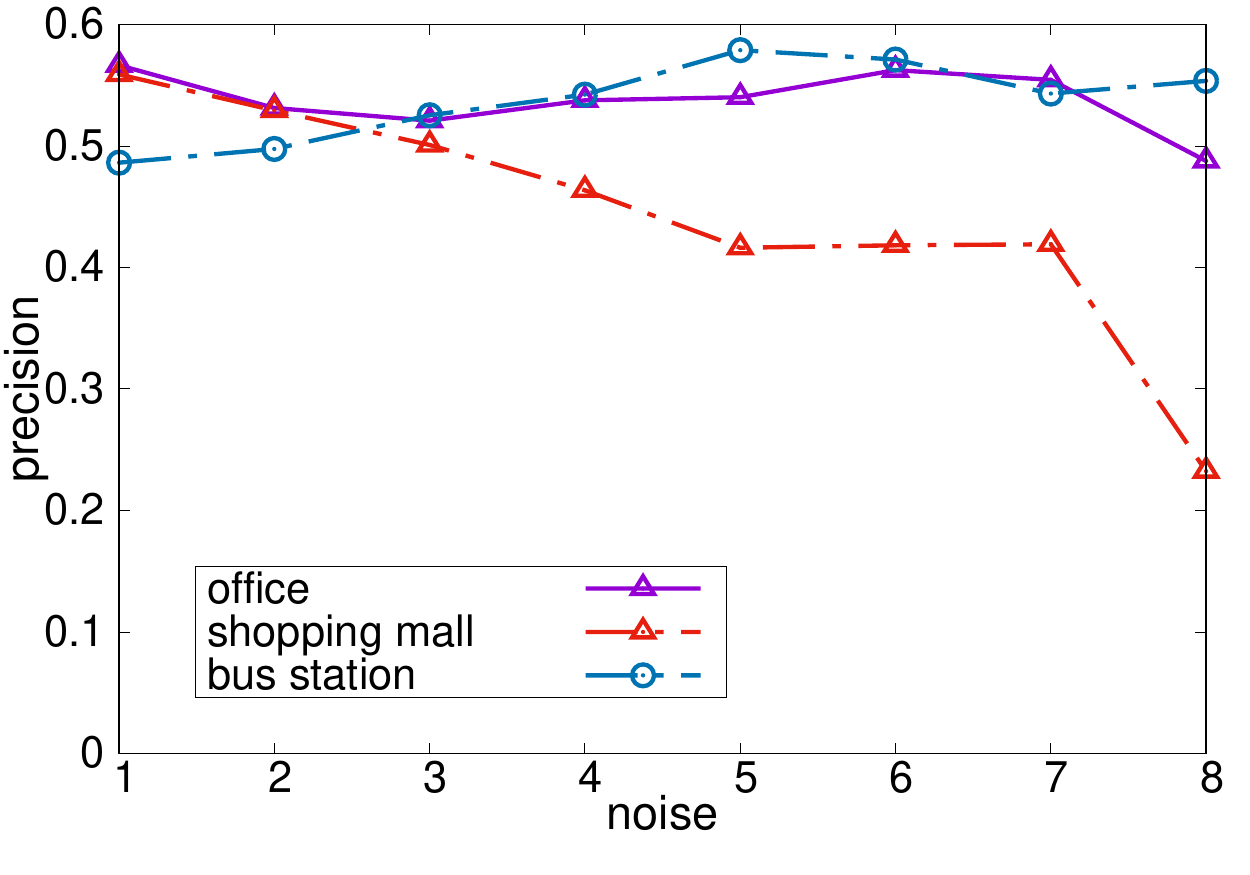}
    }
    
    \subfigure[\small Recall.]{
        \label{fig:noise_recall}
        \includegraphics[width=0.33\textwidth]{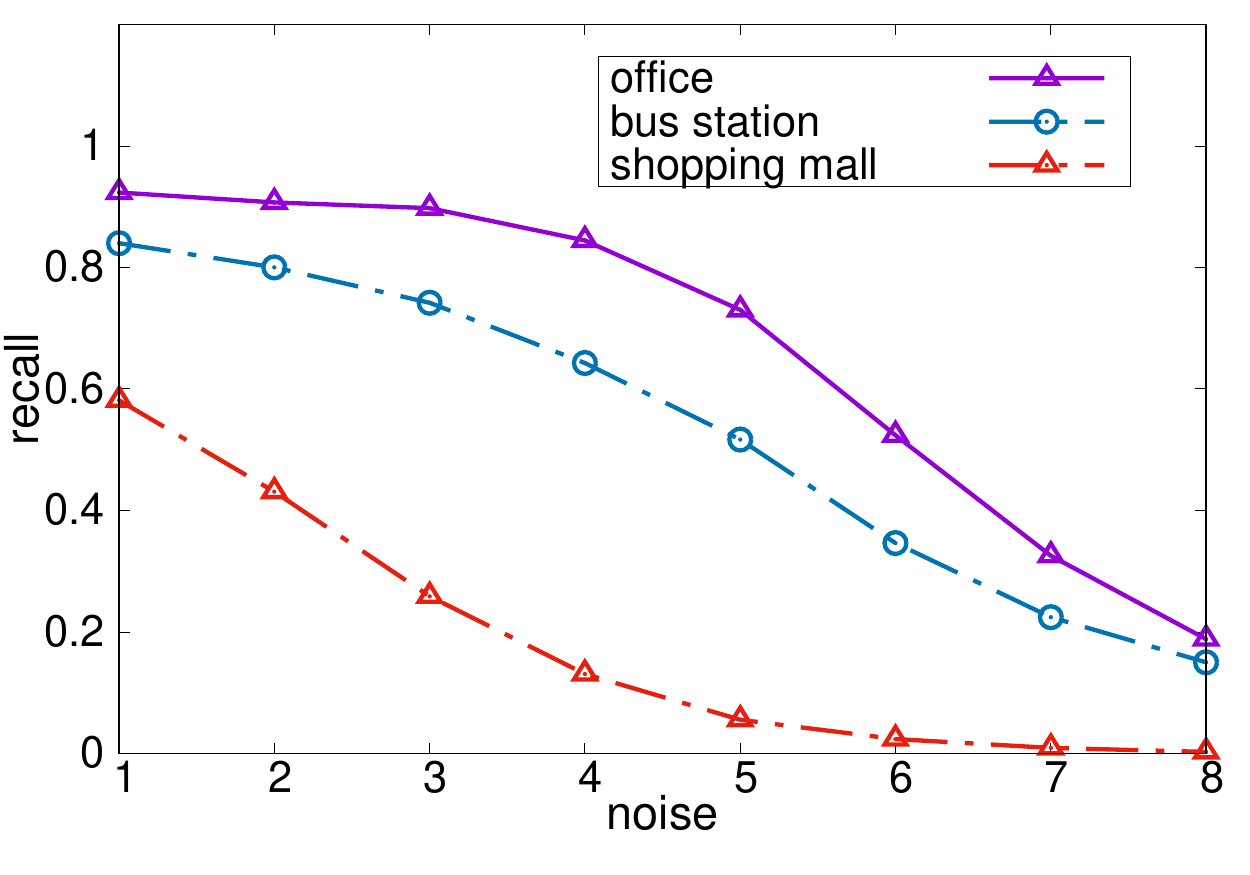}
    }
     \vspace{.6in}
    \subfigure[\small F1-score.]{
        \label{fig:nosie_f1}
        \includegraphics[width=0.33\textwidth]{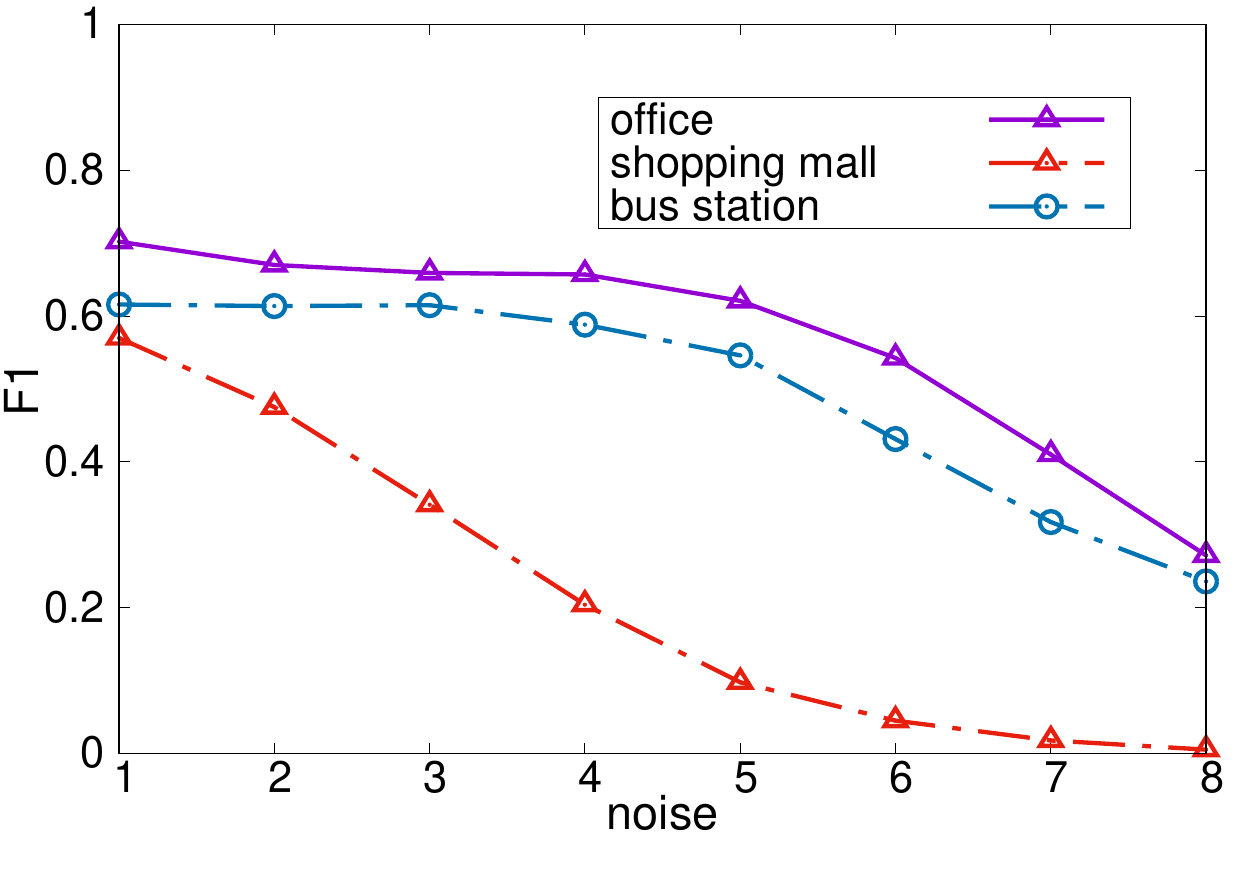}
    }
}
\caption{Impact of different noise levels on the performance of contact detection.}
\label{fig:impact_noise}
\end{figure*}

In this part, we evaluate the impact of AP number on the performance when the contact proximity is set as $k = 2$m. . 
We randomly filter $\sigma\%$ signals from the signal vectors for each site, and compare the signal similarity of two devices for contact detection. The filtering rate $\sigma\%$ is set to be $10\%$ -- $90\%$. The precision and recall versus the average signal number are presented in Figure \ref{fig:impact_ap_num}. 

In Figure \ref{fig:impact_ap_num}, as the average signal number increases, the precision increases slightly. The precision is still acceptable when the average signal number is small. Even removing $90\%$ of the signals, the precision does not drop significantly for the office and shopping mall sites. The precision outdoors~(the bus station) is more stable than others. The recall shown in Figure \ref{fig:impact_ap_num} does not have obvious change as the signal number changes, demonstrating the robustness of our approach.

\subsection{Environmental dynamicity}
\label{sec:dynamic environment}

APs in a site may change at different times, for example, some APs may shut down or the RSSIs may be different. We study the impact of the difference of APs and RSSIs on the performance of \sysname{}. Following the previous experiments, two phones are put at a distance of $2$m for data collection.

\begin{figure*}
\noindent\makebox[\linewidth][c]{
    \subfigure[\small Precision.]{
        \label{fig:different_device_precision}
        \includegraphics[width=0.33\textwidth]{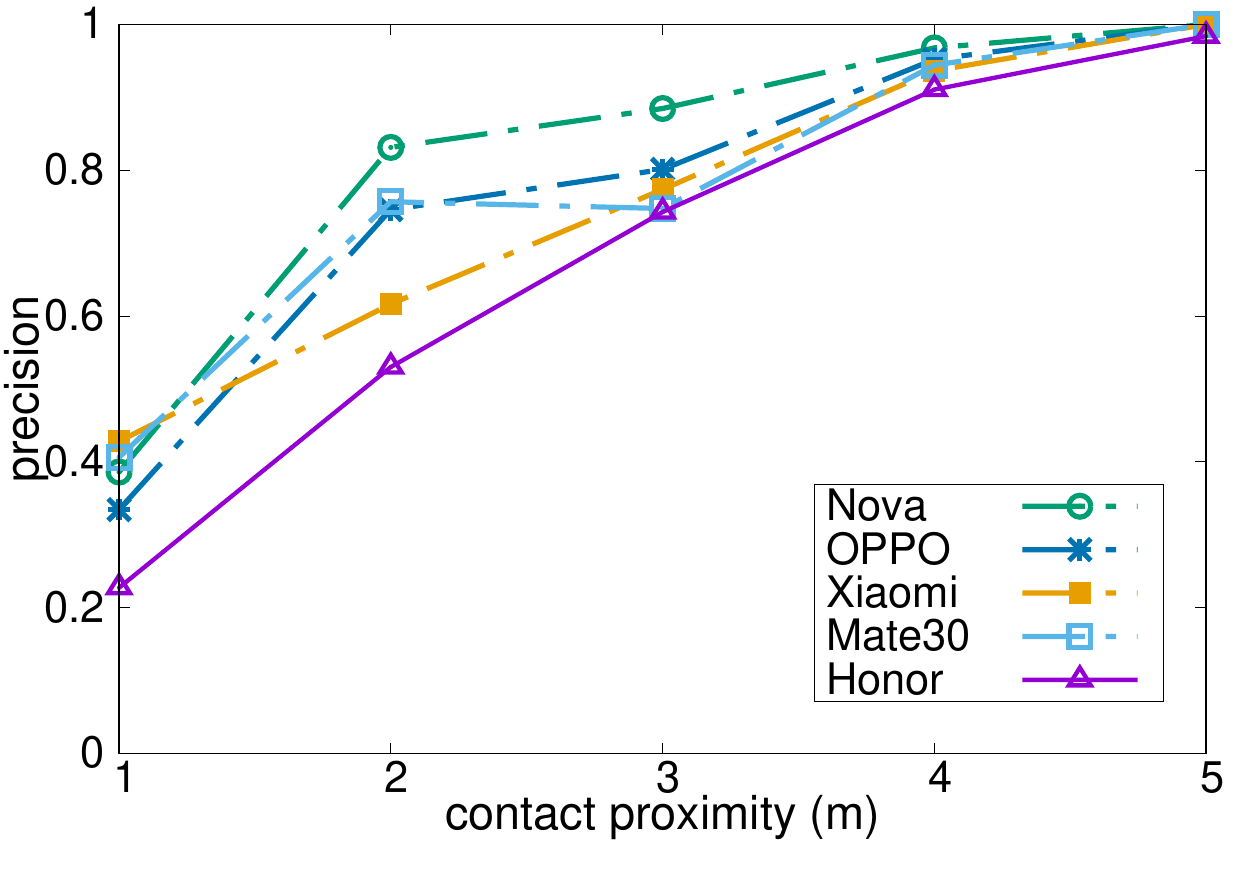}
    }
    \vspace{.6in}
    \subfigure[\small Recall.]{
        \label{fig:different_device_recall}
        \includegraphics[width=0.33\textwidth]{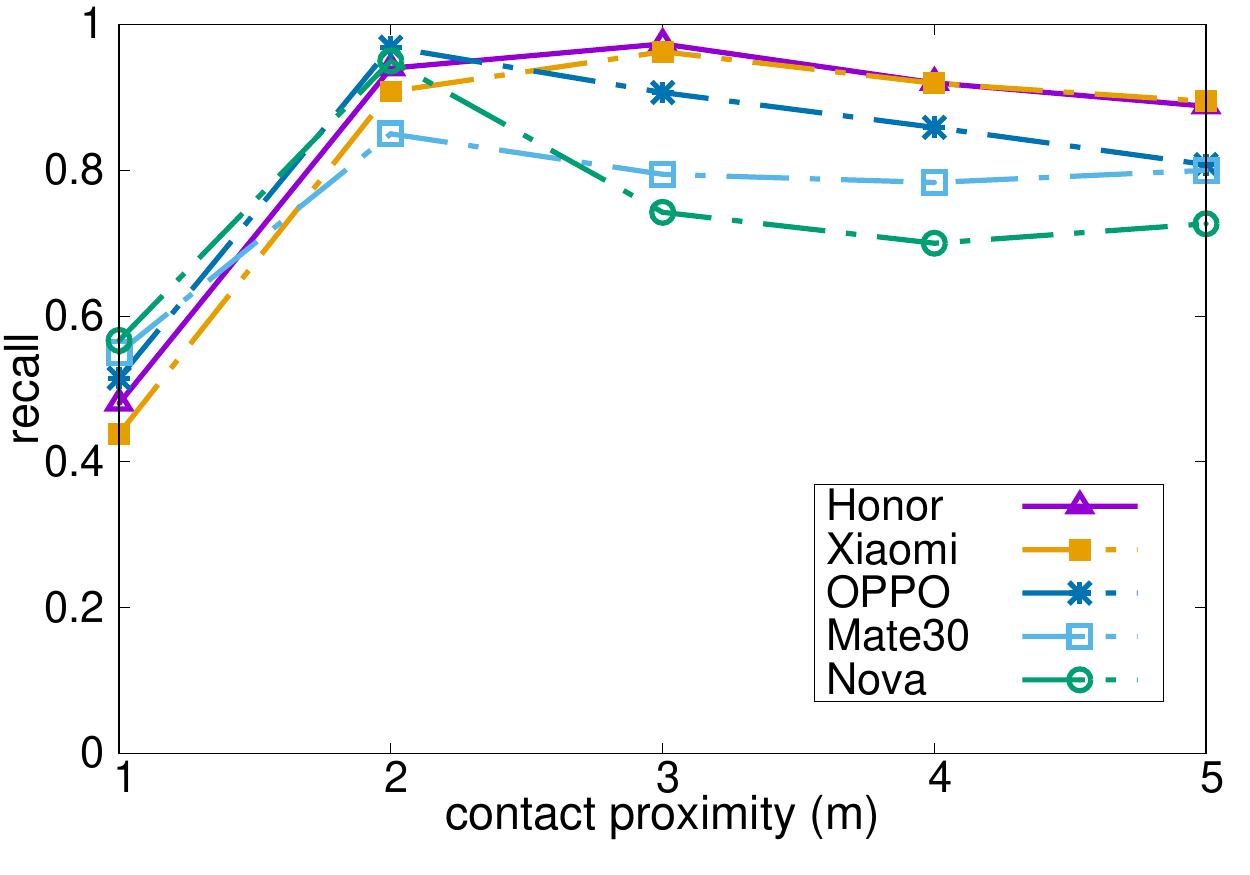}
    }
    \vspace{.6in}
    \subfigure[\small F1-score.]{
        \label{fig:different_device_f1}
        \includegraphics[width=0.33\textwidth]{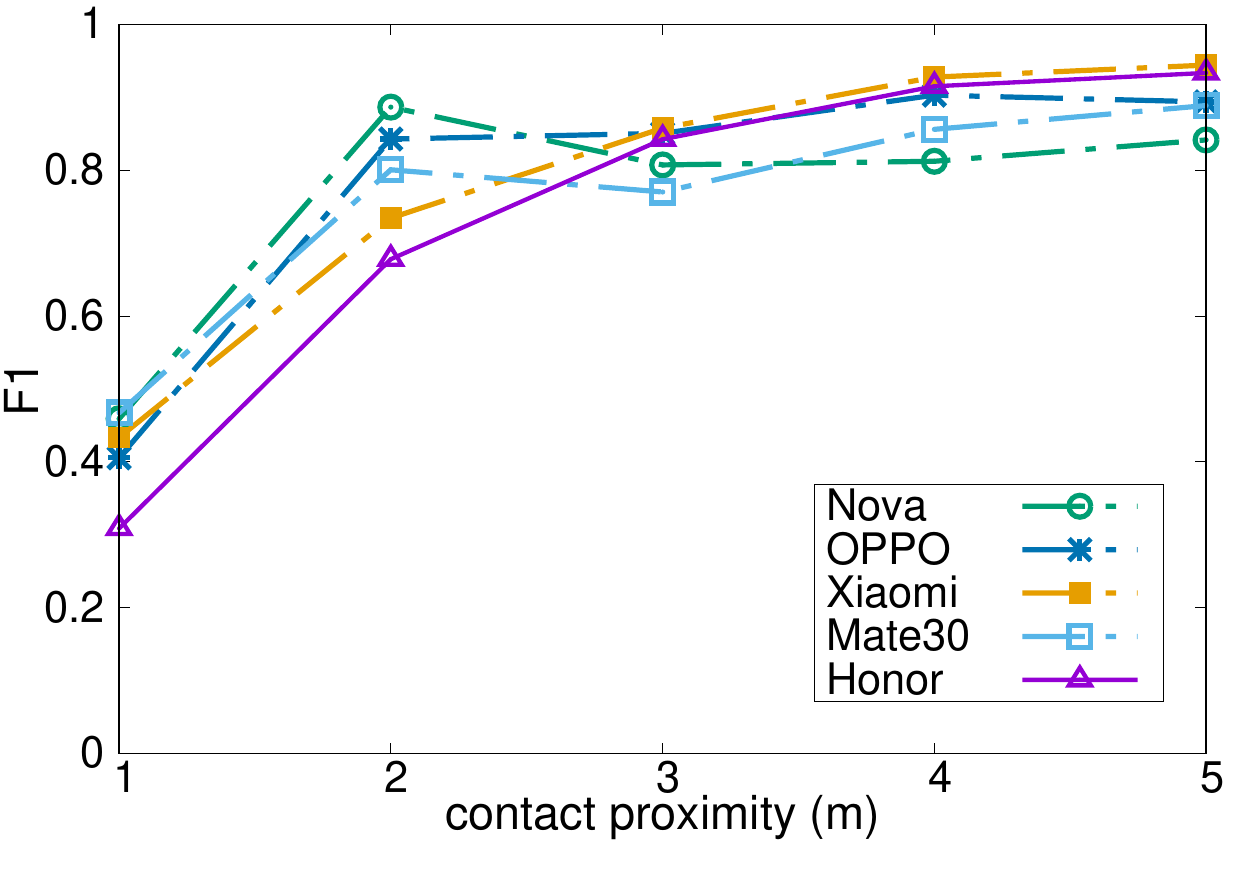}
    }
}
\caption{Performance of different devices.}
\end{figure*}

To study the impact of the difference of APs, we filter out $\sigma\%$ signal IDs from the signal profile of a phone while another remain the same. The filtering rate $\sigma\%$ is set to be $10\%$ -- $90\%$. The precision, recall and F1 score versus the filtering rates are presented in Figure \ref{fig:impact_filtering_rate}. As shown in Figure \ref{fig:filtering_rate_precision}, when more signals are filtered~(i.e., $\sigma\%$ becomes larger), the precision, recall and F1 score all decline. However, even with $50\%$ of the signals filtered, \sysname{} still achieves good performance in the three sites, which illustrates the robustness of \sysname{} w.r.t the difference of APs.

Furthermore, to evaluate the impact of the difference of RSSIs, we add a Gaussian noise to the RSSIs in one phone's signal profile as follows,
\begin{equation}
     s_i = s_i +  d, d \sim \text{Gaussian}(0, \gamma),\\
\end{equation}
where $s_i$ is the raw RSSI and d is the Gaussian noise. $\gamma\%$ is set to be $1$ -- $8$. The precision, recall and F1 score versus the filtering rates are presented in Figure \ref{fig:impact_noise}. 
The precision of \sysname{} at the bus stop and shopping mall increases slightly when the noise becomes larger. The reason is that when the noise becomes large, the signal similarity becomes smaller and false positive declines. However, the recall and F1 score drop with the increase in the noise. Overall, the performance of \sysname{} remains good when the noise is small~(less than $3$) but it drops significantly when the noise is large.

\subsection{Heterogeneous devices}
\label{sec:heterogeneous_devices}

Different devices have different abilities to scan WiFi signals. Two co-located devices may scan different signals and RSSIs. We evaluate the performance of different devices. For each device, we compare its data at $\ell_0$ with other devices' data at $\ell_i$~($i > 0$) in the same site. We set the contact proximity as $1$m -- $5$m, and set the threshold following the discussion in Section \ref{sec:threshold}. Precision, recall and F1-score are used as metrics.

The  precision  versus contact proximity for different devices in the office site is presented in Figure \ref{fig:different_device_precision}. Given the contact proximity, the precision is different for distinct devices, which is consistent with our discussion. As the contact proximity increases, the precision of all devices increases. The precision of all devices significantly increases when $k \geq 2$m. 
The recall versus contact proximity for different devices in the office is presented in Figure \ref{fig:different_device_recall}. Similar to the result of precision, the performance of all devices has a large improvement in recall when $k = 2$m.  All devices achieve high recall when $k \geq 2$m, indicating the good performance of our approach on recall. The F1-score result is shown in \ref{fig:different_device_f1}, which demonstrates the good overall performance of all tested devices.
The results demonstrate that our approach is effective and can be applied to phones of different brands. 

\subsection{Data sampling rate}
\label{sec: data_sampling_rate}
Since the APs and their RSSIs of a site do not change in a short time, the impact of data sampling rates is not obvious when users are stationary. Consequently, we discuss its impact for the scenario when users are moving. 

\begin{figure}
    \centering
  \includegraphics[width=.65\linewidth]{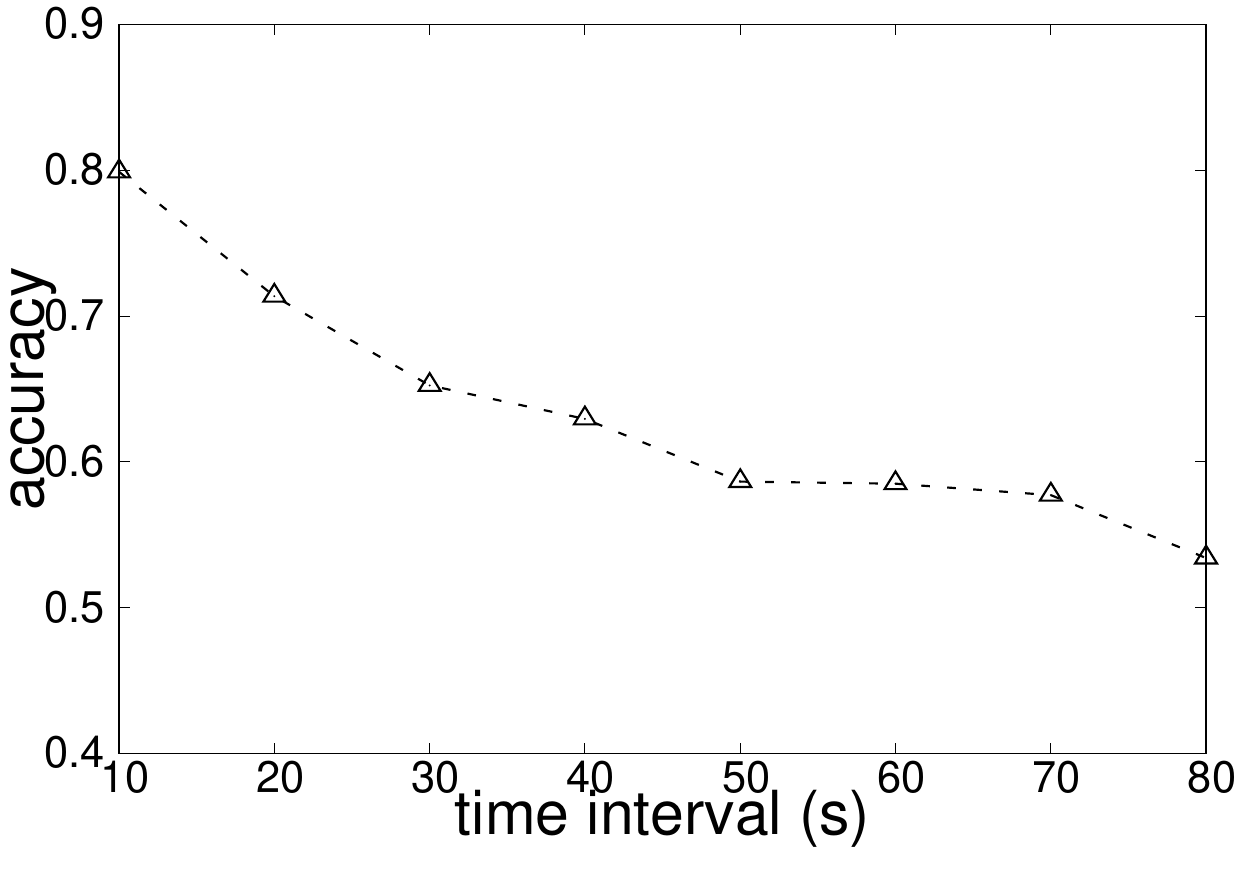}
  \caption{Impact of different time intervals. }
  \label{fig:time_interval}
\end{figure}

Some users are walking in groups in the campus with their mobile phone to collect WiFi data. The time interval of data is set as $10$s -- $80$s in the experiment. For the data of the same group, we use recall as the metric to evaluate the performance of contact detection. The result of recall is presented in Figure \ref{fig:time_interval}. As the time interval becomes larger, the recall of the detection declines. 
The reason is that users may scan WiFi data at two locations where the similarity of the WiFi is significantly different when users are moving and the time interval is larger. As a result, contact is more difficult to detect.

%% file: 7_case_study.tex
\section{IoT Implementation as a Case Study}

With the \sysname{} SDK, we have installed it into Android smart watches whose data are synced to one's phone.  Through an app, the user is notified of his/her exposure duration to the virus.
We report the smart watch implementation details and user interface
in Section~\ref{sec:details}.
Besides smart watches, we have also installed the SDK on Android phones.
We validate its design and 
 performance in Section~\ref{sec:test}.

\subsection{Smart watch implementation}
\label{sec:details}
\begin{figure}
  \centering
  \includegraphics[width=.99\linewidth]{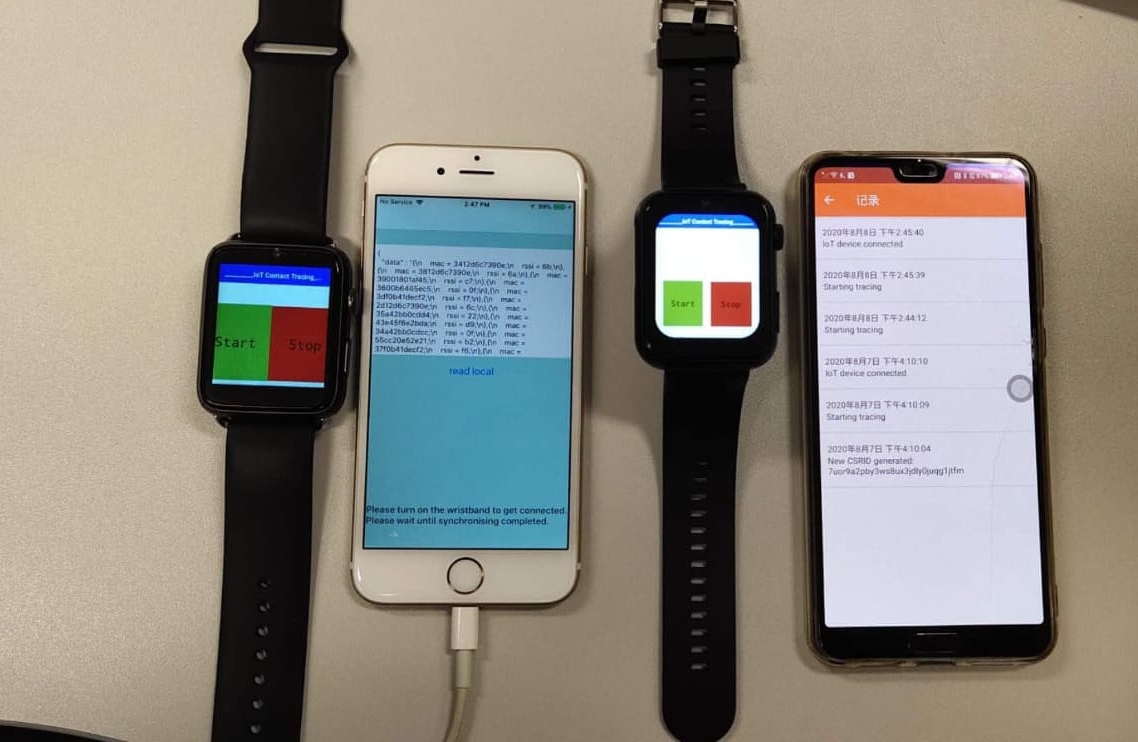}
  \caption{Data transition from smart watches to Android and IOS phones.}
  \label{fig:data_transition}
\end{figure}

Our SDK can be run independently on Android phones for data collection and exposure detection. It can also run on Android IoT devices to collect data and transit the data to an Android/iOS phone for exposure detection.  We show in Figure~\ref{fig:data_transition} an IoT smart watch system we have built based on \sysname{}.   The smart watches pair with phones through an Android or iOS app, as illustrated here with an Android phone and iPhone.  Scanned data of the smart watches are synced with their phone apps for exposure computation and notification.

\label{Sec:case study}

\begin{figure*}[ht]
\noindent\makebox[\linewidth][c]{
    \subfigure[\small Exposure data collection.]{
        \label{fig:Exposure data collection}
        \includegraphics[width=0.25\textwidth]{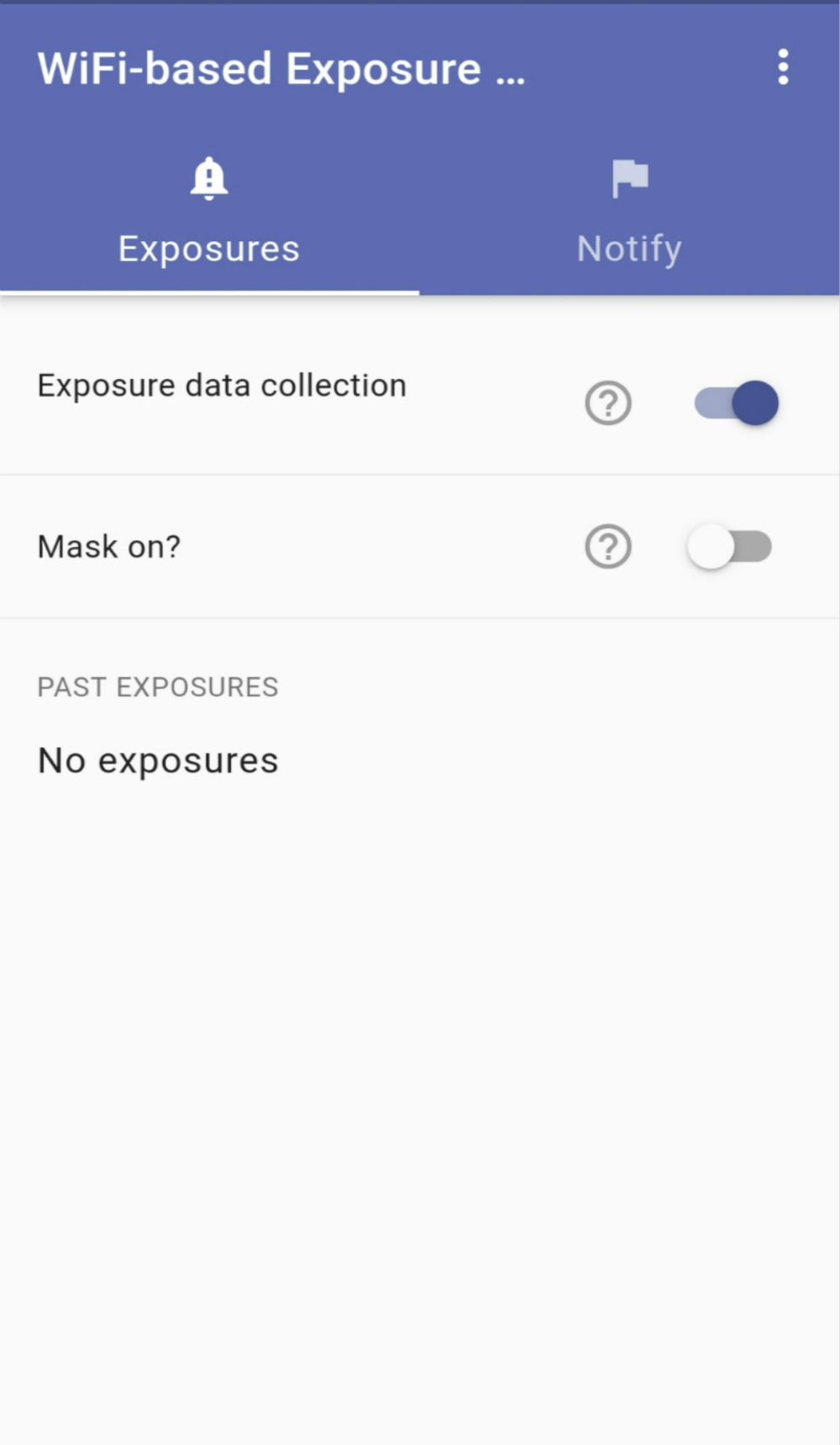}
    }
    \vspace{.6in}
    \subfigure[\small Share positive test result.]{
        \label{fig:Share positive test result}
        \includegraphics[width=0.25\textwidth]{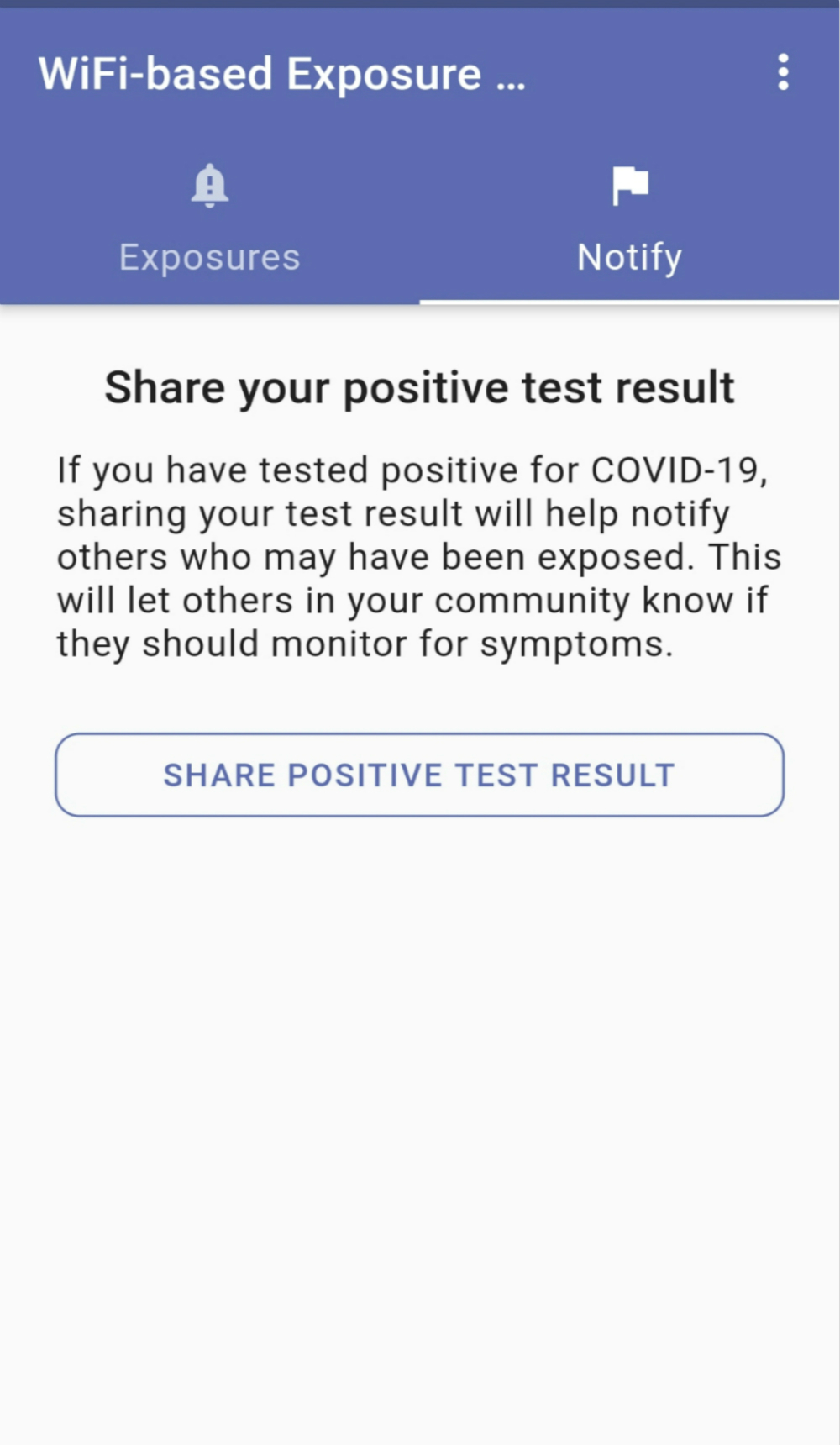}
    }
     \vspace{.6in}
    \subfigure[\small Possible exposure notification.]{
        \label{fig:Possible exposure notification}
        \includegraphics[width=0.25\textwidth]{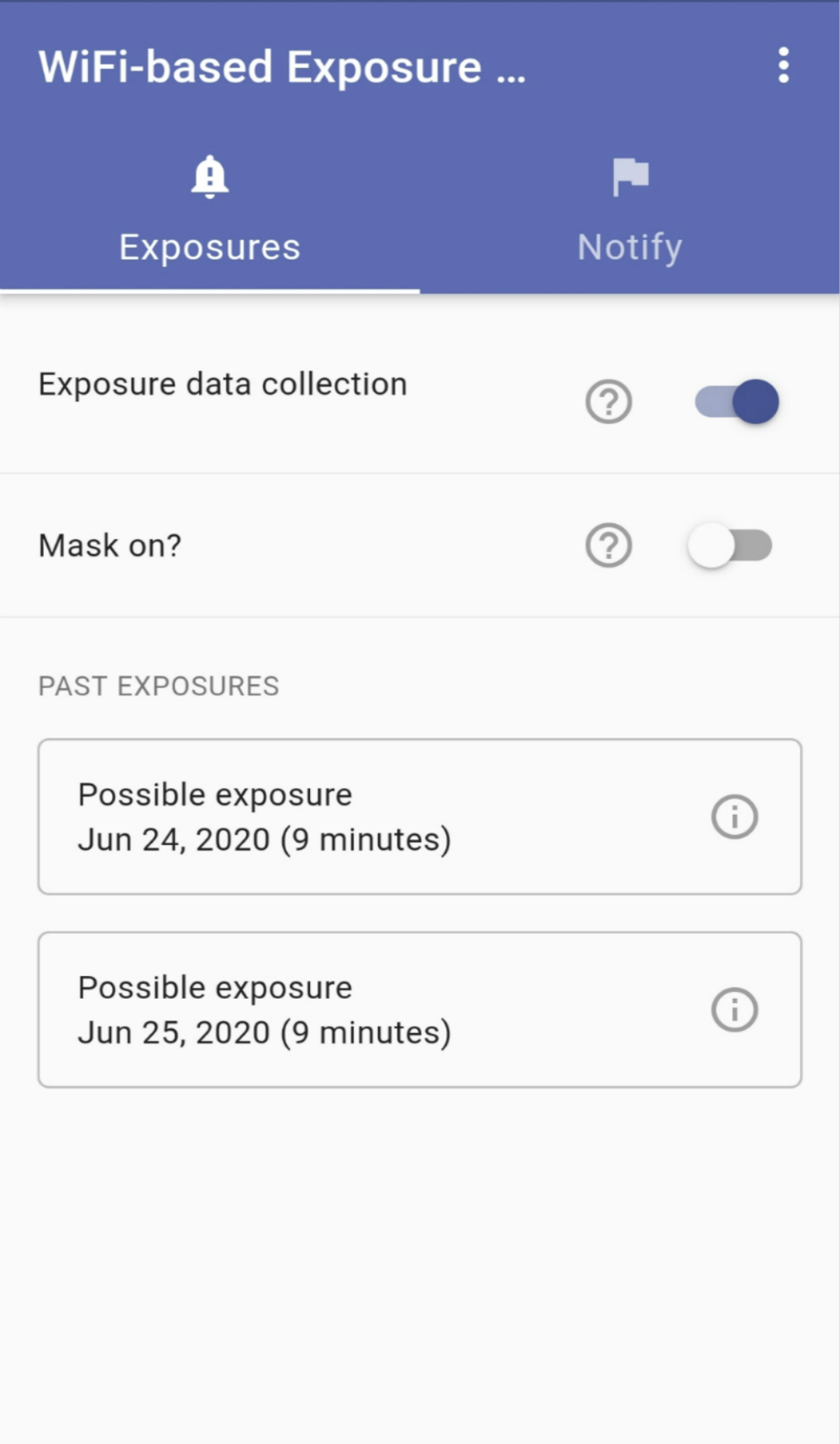}
    }
    \vspace{.6in}
    \subfigure[\small Testing mode.]{
        \label{fig:testing_mode}
        \includegraphics[width=0.25\textwidth]{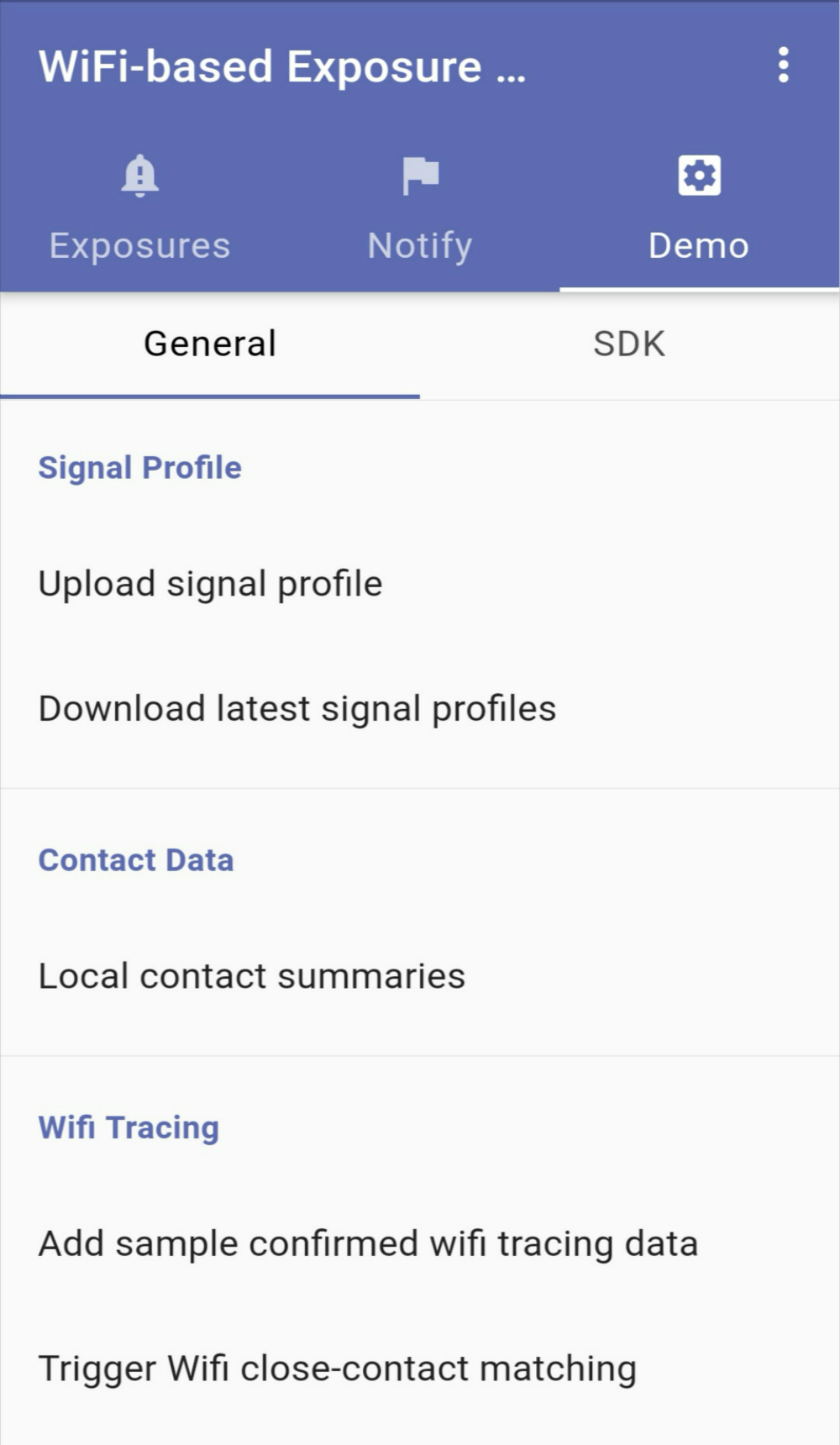}
    }
}
\caption{The user interface of an app to notify users of viral exposure duration.}
\label{fig:case_study}
\end{figure*}

The user interface of the phone app is shown in Figure~\ref{fig:case_study}. As shown in Figure~\ref{fig:Exposure data collection}, once a user turns on the button of ``Exposure data collection'', the app will start to scan nearby WiFi and store the data locally every 1 minute. Users may turn off data collection anytime and anywhere for personal reasons. The signal IDs~(i.e., the AP MAC addresses) are encrypted when the data are stored. If a user is confirmed as being infected, she/he could upload her/his signal profile to the server~(Figure \ref{fig:Share positive test result}), so that others could download the data for matching. If a user has close contact with a confirmed case, she/he will receive a notification, showing when the close contact happened and how long the contact duration was~(Figure \ref{fig:Possible exposure notification}). In the app, data are downloaded and matched automatically every day. For the purpose of testing, we also have a testing mode as shown in Figure~\ref{fig:testing_mode}, by which we can download the data, and trigger the detection manually during the testing.

\subsection{Testing and validation for phones}
\label{sec:test}

Besides smart watches, we have installed the SDK on five Android phones, namely Honor, Huawei Nova, Huawei Mate30, Xiaomi, and OPPO.  We present here a case study on these phones.
We set the contact proximity as $2$m for testing. The app collects WiFi data every $1$ minute. Hence, the detection approach introduced in Section \ref{sec:contact_detection} will report a detection result~(i.e., true or false) for the data at each minute. In our testing, if a user stays with the virus within $2$m for more than $5$ minutes in a $10$-minutes sliding time window, she/he will receive a possible exposure notification. The virus lifespan is set to be $30$ minutes. Note that the contact duration, the length of the sliding time window and the virus lifespan are parameters for the app, which can be changed according to the advice of the health officer.

We test the app in an office using the five phones.  The procedures are as follows. One of the phones is selected as the confirmed case, and other phones are put at a location which is $2$m away from the confirmed case. The button ``Exposure data collection'' is turned on for $15$ minutes. Then, the confirmed case uploads its signal profile, and the other phones download the signal profile for matching. After that, we put other phones at a location which is $4$m away from the confirmed case and repeat the testing. Each phone is selected as the confirmed case in turn.  The ideal result is that a phone only receives a notification when it is $2$m away from the confirmed case but there is no notification for $4$m. The testing results are presented in Tables \ref{table: exposure notification within 2m} and \ref{table:  exposure notification within 4m}. A $\surd$ represents that a phone receives a notification, while $\times$ means it does not receive a notification.

Table \ref{table: exposure notification within 2m} shows the results of exposure notification for $2m$.  It illustrates the good performance of our app for exposure notification. The performance of the Honor phone is not as good as that of other phones, indicating the different ability of phones to scan WiFi signals.

We show the results of exposure notification for $4$m in Table  \ref{table: exposure notification within 4m}. Compared with the results in Table \ref{table: exposure notification within 2m}, more phones are detected as having non-close contact, which is consistent with our expectation. Performance is distinct for different phones, but the overall performance is good.

\begin{table*}
\caption{Result of exposure notification for a separation of $2$m.}
\label{table: exposure notification within 2m}
\centering
\begin{tabular}{|c||c|c|c|c|c|}
\hline
\diagbox{\textbf{Confirmed Case}}{\textbf{User}}& Honor & Mate 30  & OPPO & Huawei Nova & Vivo  \\
\hline
\hline
Honor                                            &  --    &       $\surd$     &    $\surd$  &   $\surd$          &  $\surd$     \\
\hline
Mate30                                           &   $\times$    &     --      &  $\surd$    &     $\surd$        &   $\surd$    \\
\hline
OPPO                                             &    $\surd$   &     $\surd$        &   --   &     $\surd$        &    $\surd$   \\
\hline
Huawei Nova                                      &    $\times$   &  $\surd$      &      $\surd$&     --        &     $\surd$  \\
\hline
Xiaomi                                             &  $\times$     &    $\surd$    &  $\surd$    &   $\surd$          &   --   \\
\hline
\end{tabular}
\end{table*}

\begin{table*}
\caption{Result of exposure notification for a separation of $4$m.}
\label{table:  exposure notification within 4m}
\centering
\begin{tabular}{|c||c|c|c|c|c|}
\hline

\diagbox{\textbf{Confirmed Case}}{\textbf{User}}& Honor & Mate 30 & OPPO & Huawei Nova & Vivo  \\
\hline
\hline
Honor                                            &   --   &       $\times$       &    $\times$  &   $\surd$          &  $\surd$     \\
\hline
Mate30                                           &   $\times$    &     --       &  $\surd$    &     $\times$        &   $\times$    \\
\hline
OPPO                                             &    $\times$   &     $\surd$       &  --    &     $\surd$        &    $\surd$   \\
\hline
Huawei Nova                                      &    $\times$   &  $\times$        &      $\times$&       --      &     $\times$  \\
\hline
Xiaomi                                             &  $\times$     &    $\surd$      &  $\surd$    &   $\surd$          &  --    \\
\hline
\end{tabular}
\end{table*}

%% file: 9_conclusion.tex
\section{Conclusion and future works}
\label{sec:conclusion}
We have proposed \sysname{}, a novel WiFi-based private IoT contact tracing scheme with virus lifespan which may be spatial-temporally different due to the sanitization process.
By detecting close contact based on the similarity of WiFi data,
\sysname{} captures both direct face-to-face 
and indirect environmental contact.
To the best of our knowledge, this is the first work to consider the virus lifespan for private contact tracing using WiFi. We propose and study data processing approaches and a signal similarity metric for close contact detection.
Due to the ubiquity of  WiFi signals, 
\sysname{} can be pervasively deployed for contact tracing.

We conduct extensive experiments on \sysname{} by implementing it as an SDK. Our experimental results show that \sysname{}
achieves high precision, recall and F1-score~(up to $90\%$  when the contact proximity is $2$m) for different experimental sites, and its performance is robust against AP numbers, and devices of different brands.
Even with a small number of signals~($5$), \sysname{} still achieves good performance. This 
mean it is widely applicable to city or suburban areas.
It is also robust against 
environmental changes to detect indirect contact,
even if a substantial fraction ($50\%$) of the APs have been changed.
We have installed \sysname{} SDK into IoT devices of smart phones and watches,
and validate the simplicity, implementability and efficiency of our design.

\input{8_discussion}

%% file: 8_discussion.tex


We discuss below the possible future directions of the work.
One is to extend \sysname{} so that it can  be integrated with other non-RF signals such as INS and geomagnetism to strengthen its contact tracing capability, especially in areas where WiFi or Bluetooth are not available.
%
%
Another direction is 
to identify those dynamic or ephemeral APs~(e.g., hotspots of smart phone) from their MAC address, so that they could be
filtered out in contact tracing.  To this end, we can build a dynamic, scalable and crowd-sourced reference database for those permanent MACs to improve further the robustness of \sysname{}.
Yet another direction is to strike a balance between power conservation  and WiFi scanning frequency.
To preserve battery without compromising on tracing accuracy,
we may use a lower data sampling rate when users are stationary and a higher one when moving. To achieve this, we need to devise a dynamic data sampling algorithm by estimating  user activity using INS or other signals.